\documentclass[lettersize,journal]{IEEEtran}

\usepackage{amsmath,amsfonts}
\usepackage{amssymb}
\usepackage{algorithmic}
\usepackage{array}
\usepackage[caption=false]{subfig}
\usepackage{textcomp}
\usepackage{stfloats}
\usepackage{url}
\usepackage{verbatim}
\def\BibTeX{{\rm B\kern-.05em{\sc i\kern-.025em b}\kern-.08em
    T\kern-.1667em\lower.7ex\hbox{E}\kern-.125emX}}
\usepackage{balance}

\usepackage{graphicx}
\graphicspath{{fig/}}

\usepackage{physics}
\newcommand{\sgn}{\mathrm{sgn}}

\usepackage{booktabs}
\usepackage{cite}
\usepackage{pdfpages}

\begin{document}
\title{Temporal Evolution of Blackbody Radiation in a One-Dimensional Photonic Time Crystal}

\author{Luis Cortes-Herrera, Naren Ganesh, Jack Hasty, and Yuzhe Xiao
\thanks{Luis Cortes-Herrera, Jack Hasty and Yuzhe Xiao are with the Department of Physics, University of North Texas, Denton, Texas, USA (e-mail: yuzhe.xiao@unt.edu)}
\thanks{Naren Ganesh is with the Texas Academy of Mathematics and Science, University of North Texas, Denton, Texas, USA}
}


\maketitle

\begin{abstract}
    One of the most intriguing phenomena in time-varying-media photonics is the amplification of light in a photonic time crystal (PTC). However, studies to date have focused exclusively on PTC-based amplification of coherent light. Here, we theoretically investigate the PTC-based amplification of thermal radiation, specifically blackbody radiation. Such amplification is not only fundamentally intriguing due to the thermal radiation's stochastic nature, but also technologically relevant because thermal radiation's ubiquity, which implies that its amplification generally accompanies that of coherent radiation. For simplicity, and due to the experimental relevance of PTC amplification in transmission lines, we consider amplification in a one-dimensional medium. To analyze the amplification of blackbody radiation in a PTC, we examine the electromagnetic fields' spatial correlations and spatial spectra. We show that the spatial spectra of initially blackbody radiation converge periodically toward Gaussian profiles with progressively increasing amplitudes, coherence lengths, and spatial- and wavenumber-domain purities. We demonstrate that these asymptotic behaviors are governed by the PTC momentum band structure and can be understood using a rotating-wave approximation for the pseudo-Hermitian dynamics of the electromagnetic field in a PTC. Beyond revealing the fundamental evolution of thermal radiation in time-varying media, our numerical framework provides a general approach for analyzing the dynamics of stochastic electromagnetic fields in PTCs and time-varying media, in general. The results also provide physical insight and practical guidance for the design of PTC-based amplifiers, where the concurrent amplification of parasitic thermal radiation may occur.
\end{abstract}


\section{Introduction}


Driven by advances in ultra-fast modulation of photonic media and the growing demand to engineer optical signals, time-varying photonics \cite{galiffi2022photonics,galiffi2025electrodynamics} has emerged as a paradigm for surpassing the limits of static photonic systems. Under this paradigm, researchers have demonstrated novel electromagnetic-wave phenomena such as frequency-shifting \cite{xiao2011spectral,fan2016integrated,wright2017spectral,zhang2019electronically,hu2021chip}, optical nonreciprocity \cite{sounas2017non}, optical buffering \cite{yanik2004stopping,yanik2005stopping,zhang2019electronically,cortes2026theory}, negative refraction \cite{pendry2008time}, synthetic light drag \cite{huidobro2019fresnel}, and amplification via photon-pair generation in photonic time crystals (PTCs) and via cascaded photon up-conversion \cite{lyubarov2022amplified,asgari2024theory,galiffi2019broadband,horsley2023quantum}. Time-varying-media photonics have also been investigated as an approach to manipulate thermal radiation beyond its limits in static media \cite{vazquez2024review,liu2022thermal}, such as breaking Lorentz reciprocity to bypass Kirchhoff’s law \cite{fernandez2021extreme,liu2022thermal,ghanekar2022violation}, manipulating radiative heat flow \cite{buddhiraju2020photonic,yu2024time}, overcoming the blackbody limit and amplifying quantum vacuum fluctuations \cite{vazquez2023incandescent}, and shaping the radiation’s spatial coherence \cite{yu2023manipulating}.

One of the most fascinating photonic phenomena in time-varying media is the amplified emission and lasing in PTCs \cite{lyubarov2022amplified,asgari2024theory}. This is the phenomenon that occurs when a coherent radiation source is embedded in a PTC, and consequently, its radiation within the PTC’s momentum bandgap is exponentially amplified in time. While this phenomenon is fundamentally intriguing, it has practical limitations. First, the amplified wavelength must fall within the bandwidth of the emitter, making the PTC source inherently non-tunable. Second, the analysis in Ref.~\cite{lyubarov2022amplified} assumes lossless media and no background thermal radiation, conditions which are hardly met in practice. For instance, transparent conducting oxides in their epsilon-near-zero regime are promising candidates to implement PTCs \cite{lyubarov2022amplified,galiffi2022photonics}, but they possess non-negligible material loss. Such lossy PTCs must emit non-vanishing thermal radiation. Moreover, even for PTCs with negligible material losses, the environment’s background thermal radiation is ubiquitous. Thus, thermal radiation from these sources within the PTC’s bandgap is generally present and amplified concurrently. Besides its aforementioned technological relevance, PTC-based amplification of thermal radiation is fundamentally intriguing because it is intrinsically stochastic and conventionally broadband \cite{mandel1995optical,baranov2019nanophotonic}. Hence, the analysis of amplification of thermal radiation in PTCs requires specialized analysis, beyond that of coherent optical radiation.



Thus, in this paper, we examine PTC-driven amplification of thermal radiation, specifically, blackbody radiation \cite{mandel1995optical}. For simplicity and for its technological relevance \cite{asgari2024theory}, we consider the simplest non-trivial case of a one-dimensional, spatially homogeneous system. We study the time-evolution of the field's spatial spectra or, equivalently, its spatial cross- and auto-correlations \cite{mandel1995optical,wolf2007introduction,goodman2015statistical}. As expected, we show that the initially thermal radiation progressively increases its energy and spatial coherence. Furthermore, we demonstrate that the spatial spectrum converges periodically to a Gaussian shape. We show that the electromagnetic field's variances grow quasi-exponentially and oscillate with the period of modulation after several modulation periods. We demonstrate that the purity of the electromagnetic field increases as a result of parametric amplification. In the spatial-frequency domain, this purity increase occurs exclusively inside the PTC's bandgaps. In the spatial domain, the purity increase occurs over the field's coherence length. We explain this behavior of the fields' spatial spectra in terms of the PTC's band-structure and based on an adaptation of the rotating-wave approximation to the pseudo-Hermitian system of an EM field in time-varying media.


The remainder of the paper is organized as follows. In Section \ref{sec:system}, we discuss the physical system examined in this work (i.e., the PTC and thermal radiation), specifically the physical origin for the initial electromagnetic, the type of temporal modulation considered, and physical realizations of the overall system. In Section \ref{sec:spatiospectral}, we introduce the analytical framework we utilize to analyze the evolution of the initially thermal field in the PTC, i.e., our spatio-spectral description of the field variances. In Section \ref{sec:analytical}, we introduce and discuss auxiliary analytical tools to contextualize our numerically obtained results. Specifically, these are the PTC's momentum band structure, and an original rotating-wave approximation for the pseudo-Hermitian dynamics of electromagnetic waves in a PTC. In Section \ref{sec:results}, we present and examine the numerical results of this paper, describing the time-evolution of the fields' spatial correlations and spatial spectra. In Section \ref{sec:conclusion}, we present the paper's conclusions.

\section{System under consideration}
\label{sec:system}

We consider a spatially one-dimensional PTC, with longitudinal coordinate $z$. This simplest non-trivial geometry is amenable to analytical (Subsection \ref{subsec:rwa}) and numerical treatment while retaining the essential physics of spatially homogeneous, time-periodic media. It is also technologically relevant, having enabled experimental time reflections \cite{moussa2023observation,jones2024time} and PTCs \cite{reyes2015observation,jones2026demonstration}. Although a dielectric waveguide can realize such a system, dielectric-permittivity modulation is generally more challenging than capacitance modulation in transmission lines \cite{segal2026before,moussa2023observation,jones2024time,reyes2015observation,jones2026demonstration}, and waveguide modal dispersion complicates both analysis and implementation-independent interpretation.

We consider blackbody radiation arising from weak coupling to a thermal environment \cite{mandel1995optical}, and we describe it as a classical (i.e., non-quantum) stochastic field (see Suppl.~Sec.~S1 for a discussion of the validity of such a classical model). We assume this coupling is negligible over the modulation times studied; its full treatment is beyond the scope of this work.

The PTC is non-dispersive, with sinusoidally varying electric impermittivity $\eta(t)=\varepsilon^{-1}(t)$:
\begin{equation}
\eta(t)=\eta_s+H(t)\Delta\eta_0\sin\Omega t,
\label{eq:eta}
\end{equation}
where $H(t)$ is the Heaviside function, $\eta_s$ is the static and time-averaged impermittivity, and $\Delta\eta_0$ and $\Omega$ are the modulation amplitude and frequency. This smooth modulation both accommodates field components both fast and slow relative to the modulation rise and decay times, unlike stepwise models \cite{asgari2024theory}, and also facilitates analytical treatment. In particular, the wavenumber-domain Maxwell equations become analogous to those of a harmonic oscillator with sinusoidally varying resonance frequency (Suppl.~Sec.~S5), enabling an accurate rotating-wave approximation (RWA) for their pseudo-Hermitian dynamics (Subsection \ref{subsec:rwa}).

We study the time evolution of the fields' spatial correlations and spectra (Section \ref{sec:spatiospectral}). This representation decomposes field variances and correlations by wavenumber through the Wiener--Khinchine theorem, directly connecting them to the PTC band structure and dynamics.

\begin{figure*}
\centering
\includegraphics[width=\linewidth]{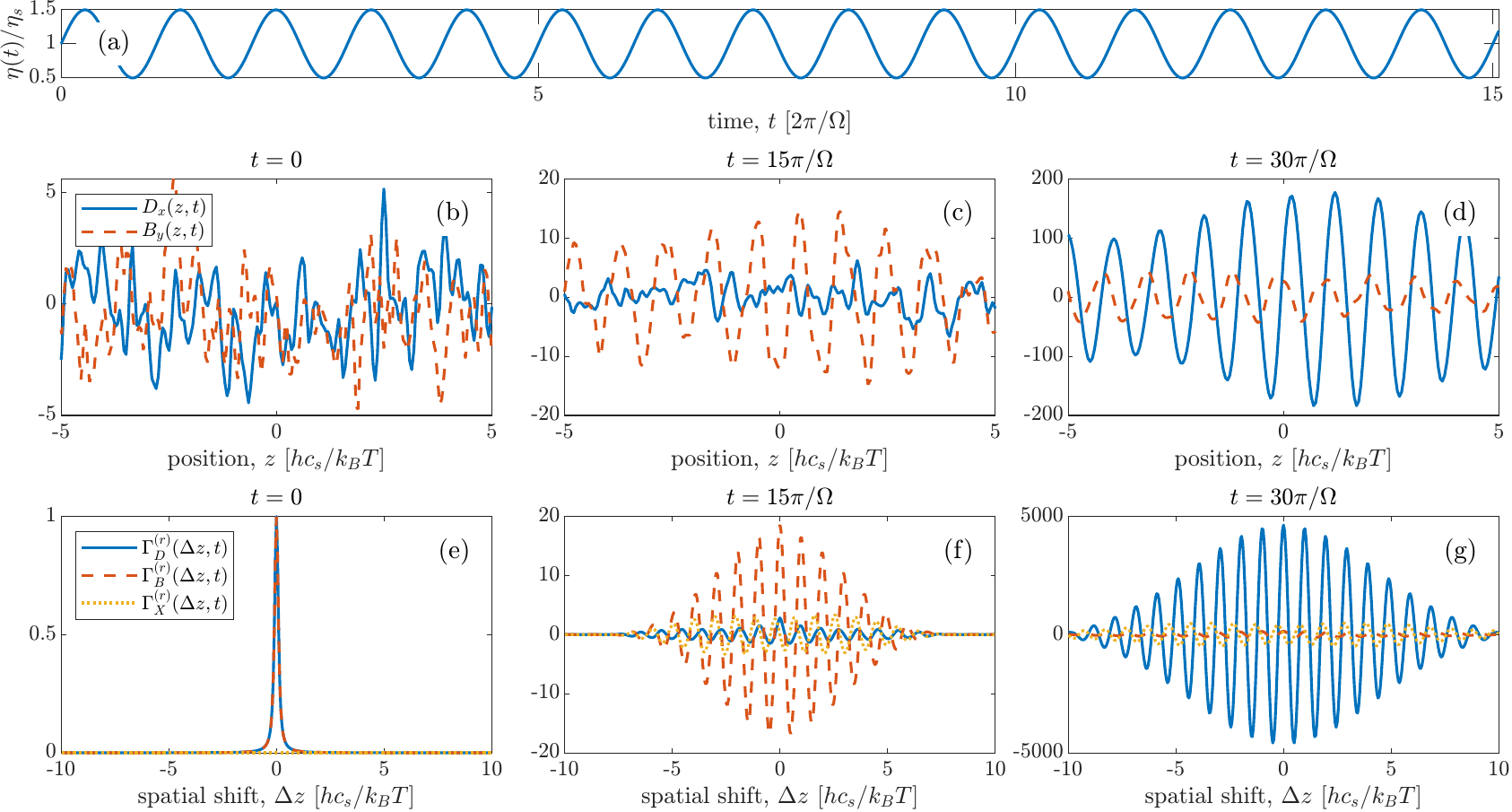}
\caption{Evolution of initially blackbody radiation in a photonic time crystal (PTC) with sinusoidally varying impermittivity $\eta(t)=\varepsilon^{-1}(t)$. (a) Spatially homogeneous impermittivity modulation at frequency $\Omega$. (b--d) Field realizations at $t=0$, $15\pi/\Omega$, and $30\pi/\Omega$. (e--g) Corresponding normalized real-valued spatial correlations: displacement auto-correlation $\Gamma^{(r)}_D$, magnetic-flux auto-correlation $\Gamma^{(r)}_B$, and displacement--magnetic-flux cross-correlation $\Gamma^{(r)}_X$.}
\label{fig:ptcblackbodycleo2edit}
\end{figure*}

Figure~\ref{fig:ptcblackbodycleo2edit} summarizes the system. Figure~\ref{fig:ptcblackbodycleo2edit}(a) shows $\eta(t)$, while Figs.~\ref{fig:ptcblackbodycleo2edit}(b)--\ref{fig:ptcblackbodycleo2edit}(d) show $D_x(z,t)$ and $H_y(z,t)$ at $t=0$, $15\pi/\Omega$, and $30\pi/\Omega$ for one stochastic realization of the initial blackbody field, calculated by finite-difference time-domain simulation (Suppl.~Sec.~S2). The corresponding spatial correlations are shown in Figs.~\ref{fig:ptcblackbodycleo2edit}(e)--\ref{fig:ptcblackbodycleo2edit}(g). With increasing modulation time, the fields are amplified, acquire longer coherence lengths, and become cross-correlated. Characterizing this evolution is the central objective of this paper.

\section{Spatio-spectral description}
\label{sec:spatiospectral}



\subsection{Definitions}
\label{subsec:definitions}

We study the second-order statistics of the displacement $D_x(z,t)$ and magnetic flux $B_y(z,t)$ rather than $E_x(z,t)$ and $H_y(z,t)$, because the former make the symmetries and conservation properties of the time-varying system more evident (Sections \ref{sec:analytical} and \ref{sec:results}). Specifically, we examine the normalized fields
\begin{equation}
\begin{pmatrix}
\tilde{D}_x(z,t) \\
\tilde{B}_y(z,t)
\end{pmatrix}
=
\frac{\sqrt{3hc_sA}}{\pi k_BT}
\begin{pmatrix}
D_x(z,t)/\sqrt{\varepsilon_s} \\
B_y(z,t)/\sqrt{\mu_0}
\end{pmatrix}.
\label{eq:dbnormal}
\end{equation}
Here, $\hbar$, is Planck's reduced constant; $k_B$, Boltzmann's constant; $T$, the initial field temperature; $c_s$, the speed of light in the static mecium; $A$, the quantization area; $\varepsilon_s$, the static medium's permittivity; and $\mu_0$, vacuum permeability. We assume a non-magnetic PTC. This normalization gives unit initial variance (Suppl.~Sec.~S3),
\begin{equation}
\expval{\tilde{D}_x^2(z,0)}
=
\expval{\tilde{B}_y^2(z,0)}
=1,
\label{eq:initialvariance}
\end{equation}
where $\expval{\cdot}$ denotes ensemble average.

Due to the PTC's spatial homogeneity, the electromagnetic fields are jointly stationary processes in space. This is analogous to how in conventional optical coherence theory of time-invariant sources and media \cite{mandel1995optical,wolf2007introduction,goodman2015statistical}, the fields are stationary processes in time. Thus, the fundamental quantities of interest in this study are the fields' spatial correlations and spatial spectra. We relate the real-valued spatial correlations $\Gamma^{(r)}_j(\Delta z,t)$ to the normalized fields \eqref{eq:dbnormal} via
\begin{equation}
    \begin{split}
        \Gamma^{(r)}_D(\Delta z,t) \equiv & \expval{\tilde{D}_x(z +\Delta z,t) \tilde{D}_x(z,t)}, \\
        \Gamma^{(r)}_B(\Delta z,t) \equiv & \expval{\tilde{B}_y(z +\Delta z,t) \tilde{B}_y(z,t)}, \\
        \Gamma^{(r)}_X(\Delta z,t) \equiv & \expval{\tilde{D}_x(z +\Delta z,t) \tilde{B}_y(z,t)}.
    \end{split}
\end{equation}
In other words, $\Gamma^{(r)}_D(\Delta z,t)$ is the real-valued displacement auto-correlation; $\Gamma^{(r)}_B(\Delta z,t)$, the real-valued magnetic auto-correlation; and $\Gamma^{(r)}_X(\Delta z,t)$, the real-valued fields' cross-correlation.

To dissect the time-evolution of the electromagnetic field, it is useful to introduce the analytic spatial correlations $\Gamma_j(\Delta z,t)$ ($j = D,B,X$). These are related to the real-valued spatial correlations $\Gamma^{(r)}_j(\Delta z,t)$ through \cite{mandel1995optical,wolf2007introduction,goodman2015statistical}
\begin{equation}
    \Gamma_j(\Delta z,t) = \Gamma^{(r)}_j(\Delta z,t) + i\Gamma^{(i)}_j(\Delta z,t). \label{eq:Gammaj}
\end{equation}
In Eq.~\eqref{eq:Gammaj}, the real-valued function $\Gamma^{(i)}_j(\Delta z,t)$ is the Hilbert transform of $\Gamma^{(r)}_j(\Delta z,t)$ with respect to $\Delta z$. Furthermore, according to the Wiener-Khinchine theorem \cite{mandel1995optical,wolf2007introduction,goodman2015statistical}, the fields' analytic (single-sided) spatial spectra $W_j(k,t)$ are each given by Fourier transform of the corresponding analytic correlation $\Gamma_j(\Delta z,t)$, i.e.,
\begin{equation}
    W_j(k,t) = \int_{-\infty}^\infty \dd{z} \Gamma_j(z,t) \exp(-ikz). \label{eq:wienerkhinchine}
\end{equation}
As a consequence of Eqs.~\eqref{eq:Gammaj} and \eqref{eq:wienerkhinchine}, $W_j(k,t)$ are indeed single-sided spectra \cite{mandel1995optical,wolf2007introduction,goodman2015statistical}, i.e., $W_j(k,t) = 0$ for $k < 0$. Similarly to Eq.~\eqref{eq:wienerkhinchine}, the double-sided spectra $W^{(r)}_j(k,t)$ are each given by the Fourier transform of the corresponding real-valued correlation $\Gamma^{(r)}_j(\Delta z,t)$.

We characterize these quantities by their central wavenumbers $\bar{k}_j(t)$, spatial bandwidths $\Delta k_j(t)$, and coherence lengths $L_j(t)$. The first two are the normalized first moment and standard deviation of $|W_j(k,t)|^2$, while $L_j(t)$ is the normalized standard deviation of $|\Gamma_j(\Delta z,t)|^2$ (see Suppl.~Sec.~S4 and Refs.~\cite{wolf1958reciprocity,mandel1962measures,mandel1995optical}). Since the $\Gamma_j$ and $W_j$ form Fourier-transform pairs, they obey the Fourier-transform limit $L_j\Delta k_j\geq 1/2$, with equality if and only if $W_j(k,t)$ is a Gaussian function of $k$ (equivalently, $\Gamma_j(z,t)$ is a Gaussian function of $z$) \cite{mandel1995optical}. As is well known \cite{papoulis1977signal,goodman2017introduction}, the uncertainty product $L_j \Delta k_j$ is a measure of the complexity of $\Gamma(\Delta z,t)$ and $W_j(k,t)$.

To fully describe the field cross-correlations, it is useful to examine the spatial correlations and spatial spectra of arbitrary linear combinations of the electromagnetic fields $\tilde{D}_x(z,t)$ and $\tilde{B}_y(z,t)$. In analogy to optical coherence theory \cite{mandel1995optical,wolf2007introduction} these are given by the Hermitian forms of the spatial cross-correlation matrix $\bar{\Gamma}^{(r)}(\Delta z,t)$ and of the cross-spectral density matrix $\bar{W}(k,t)$. These are the Hermitian, non-negative-definite matrices \cite{mandel1995optical,wolf2007introduction}
\begin{equation}
    \begin{split}
        \bar{\Gamma}^{(r)}(\Delta z,t) \equiv & \begin{bmatrix}
    \Gamma^{(r)}_D(0,t) & \Gamma^{(r)}_X(\Delta z,t) \\
    \Gamma^{(r)}_X(\Delta z,t) & \Gamma^{(r)}_B(0,t)
    \end{bmatrix}, \\
    \bar{W}(k,t) \equiv & \begin{bmatrix}
        W_D(k,t) & W_X(k,t) \\
        W^*_X(k,t) & W_B(k,t)
    \end{bmatrix}.
    \end{split} \label{eq:GWmats}
\end{equation}

To dissect the Hermitian forms induced by the matrices $\bar{\Gamma}^{(r)}(\Delta z,t)$ and $\bar{W}(k,t)$, we examine their eigenvalue decomposition. Thus, we introduce a ''spatial purity" $P_\Gamma(\Delta z,t)$ and a ''spectral purity" $P_W(k,t)$, analogs of the degree of polarization from the theory of partial polarization \cite{mandel1995optical,wolf2007introduction}. These are defined in terms of the eigenvalues $\gamma_\pm(\Delta z,t)$ and $w_\pm(k,t)$ of $\bar{\Gamma}(\Delta z,t)$ and $\bar{W}(k,t)$, respectively, as
\begin{equation}
    \begin{split}
        P_\Gamma(\Delta z,t) = & \frac{\gamma_+(\Delta z,t) - \gamma_-(\Delta z,t)}{\gamma_+(\Delta z,t) + \gamma_-(\Delta z,t)}, \\
        P_W(k,t) = & \frac{w_+(k,t) - w_-(k,t)}{w_+(k,t) + w_-(k,t)}.
    \end{split}
\end{equation}
We define $\gamma_\pm$ and $w_\pm$ so $\gamma_+ > \gamma_-$ and $w_+ > w_-$. Hence, $0 \leq P_\Gamma \leq 1$ and $0 \leq P_W \leq 1$. As the degree of polarization \cite{mandel1995optical,wolf2007introduction} the spatial $P_\Gamma$ and spectral $P_W$ purities measure the fraction of the spatial-domain and wavenumber-domain field variance in a single linear combination (analogous to a single polarization) of electromagnetic fields.

The eigenvectors of $\bar{\Gamma}^{(r)}$ and $\bar W$ are normalized two-component complex vectors and may therefore be represented by Bloch-vector components \cite{damask2005polarization}. Because each Hermitian matrix has orthogonal eigenvectors whose Bloch vectors are mirror symmetric \cite{damask2005polarization,cortes2023theory}, only the eigenvector associated with the larger eigenvalue is required. We denote its components by $S_j^{(\Gamma)}(\Delta z,t)$ and $S_j^{(W)}(k,t)$ ($j=1,2,3$). Their physical interpretation and explicit forms are given in the Supplemental Document (Suppl.~Sec.~S8).

\subsection{Initial conditions}
\label{subsec:initial}

The initial state of the electromagnetic field (before modulation) is that of blackbody radiation of a one-dimensional system. Thus, the initial real-valued spatial correlations $\Gamma^{(r)}_j(z,0)$ and two-sided spatial spectra $W^{(r)}_j(k,0)$ are given by (Suppl.~Sec.~S3)
\begin{equation}
    \begin{split}
        \Gamma^{(r)}_D(z,0) = & \Gamma^{(r)}_B(z,0) = \Gamma_0(z), \\
        W_D^{(r)}(k,0) = & W^{(r)}_B(k,0)  = W_0(k), \\
        \Gamma^{(r)}_X(z,0) = & W_X^{(r)}(k,0) = 0.
    \end{split} \label{eq:initial}
\end{equation}
In Eq.~\eqref{eq:initial}, the functions $\Gamma_0(z)$ and $W_0(k)$ are given by
\begin{equation}
    \begin{split}
        \Gamma_0(\Delta z) & =  f(\pi k_B T \Delta z/\hbar c_s), \\
        W_0(k) & = \frac{6}{\pi} \left(\frac{\hbar c_s}{k_B T}\right)^2 |k| \left[\exp(\hbar c_s|k|/k_B T) - 1\right]^{-1},
    \end{split} \label{eq:initial2}
\end{equation}
with $f(x)$ the real-analytic function \cite{bourret1960coherence,gardiner2004quantum}
\begin{equation}
    f(x) = \left\{\begin{matrix}
        3 x^{-2} - 3 \csch^2x, & x \neq 0, \\
        1, & x = 0.
    \end{matrix}\right.
\end{equation}
We compute the analytic correlation functions $\Gamma_j(z)$ from the real-valued ones $\Gamma^{(r)}_j(z)$ via numerical Hilbert transform \cite{marple1999computing}.

It is intuitively clear that Eqs.~\eqref{eq:initial} and \eqref{eq:initial2} yield the correct initial spatial spectral densities $W_D(k,0)$ and $W_B(k,0)$ of $W_0(k)$. This is because, for each $k$, $W_0(k)$ in Eq.~\eqref{eq:initial2} is proportional to the energy of an optical mode of frequency $c_s |k|$, i.e., $\hbar c_s |k| [\exp(\hbar c_s |k|/k_B T) - 1]^{-1}$.

\begin{figure}
\centering
\includegraphics[width=\linewidth]{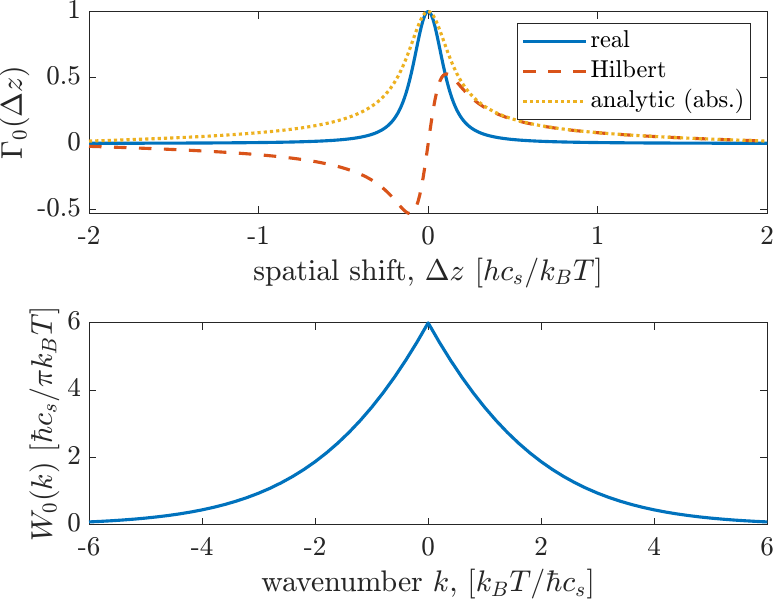}
\caption{(a) Initial one-dimensional blackbody spatial auto-correlation $\Gamma_0(\Delta z)$, its Hilbert transform, and the magnitude of the corresponding analytic auto-correlation. (b) Initial spatial spectrum $W_0(k)$.}
\label{fig:ptcblackbodyinitial}
\end{figure}

Figure~\ref{fig:ptcblackbodyinitial} shows $\Gamma_0(\Delta z)$, its Hilbert transform, the magnitude of its analytic correlation, and $W_0(k)$. The corresponding figures of merit are reported in Suppl.~Sec.~S4. Both $\Gamma_0^{(r)}$ and $|\Gamma_0|$ decay monotonically over a characteristic initial coherence length $0.2815\times hc_s/k_BT$, whereas $W_0^{(r)}$ is centered at $k=0$ and decays with $|k|$ over a bandwidth $0.7564\times k_BT/\hbar c_s$.

\subsection{Time evolution}
\label{subsec:timeevol}

Spatial homogeneity implies that only waves with identical spatial dependence $\exp(ikz)$ couple. For each $k$, these are forward- and backward-propagating waves with amplitudes $a(k,t)$ and $a^*(-k,t)$. Furthermore, the time-evolution is linear in the fields, so the modal amplitudes $a(k,t)$ evolve according to
\begin{equation}
\begin{bmatrix}
a(k,t) \\
a^*(-k,t)
\end{bmatrix}
=
U(k,t)
\begin{bmatrix}
a(k,0) \\
a^*(-k,0)
\end{bmatrix},
\label{eq:Ukt}
\end{equation}
where $U(k,t)$ is the wavenumber-domain time-evolution matrix, derived from Maxwell's equations in a PTC. The matrix $U(k,t)$  can be evaluated fully numerically or analytically within the rotating-wave approximation (Suppl.~Sec.~S6).

Naturally, Eq.~\eqref{eq:Ukt} for the field-mode amplitudes $a(k,t)$ leads to a time-evolution of the spatial spectra $W_j(k,t)$ and thus the spatial correlations $\Gamma_j(\Delta z,t)$ $(j = D,B,X)$. Due to the PTC's spatial homogeneity, and the simple form of the initial conditions \eqref{eq:initial}, each $W_j(k,t)$ for each wavenumber $k$ evolves separately as (Suppl.~Sec.~S5)
\begin{equation}
    \begin{split}
        W_D(k,t) = & W_0(k) \left[S_0(k,t) + S_1(k,t)\right], \\
        W_B(k,t) = & W_0(k) \left[S_0(k,t) - S_1(k,t)\right], \\
        W_X(k,t) = & -i\sgn(k) W_0(k) S_2(k,t).
    \end{split} \label{eq:Wjtime}
\end{equation}
In Eq.~\eqref{eq:Wjtime}, $S_j(k,t)$ are the Bloch-vector components associated with the column vector $[u_{11}(k,t), u_{21}(k,t)]^T$, i.e.,
\begin{equation}
    S_j(k,t) = \begin{pmatrix}
        u^*_{11}(k,t) & u^*_{21}(k,t)
    \end{pmatrix} \sigma_j \begin{pmatrix}
        u_{11}(k,t) \\
        u_{21}(k,t)
    \end{pmatrix}, \label{eq:Sk}
\end{equation}
In Eq.~\eqref{eq:Sk}, $\sigma_j$ ($j = 0,1,2,3$) are the Pauli matrices
\begin{equation}
    \begin{split}
        \sigma_0 = & \begin{pmatrix}
        1 & 0 \\
        0 & 1
    \end{pmatrix}, \quad
    \sigma_1 = \begin{pmatrix}
        0 & 1 \\
        1 & 0
    \end{pmatrix}, \\
    \sigma_2 = & \begin{pmatrix}
        0 & -i \\
        i & 0
    \end{pmatrix}, \quad
    \sigma_3 = \begin{pmatrix}
        1 & 0 \\
        0 & -1
    \end{pmatrix}; \label{eq:pauli}
    \end{split}
\end{equation}
and $u_{nm}(k,t)$ the $(n,m)$-th element of the time-evolution matrix $U(k,t)$ from Eq.~\eqref{eq:Ukt}.


\section{Analytical tools}
\label{sec:analytical}

The preceding framework fully specifies the evolution of initially blackbody radiation in a one-dimensional PTC. To connect this evolution to the underlying physics, we now introduce two analytical tools: the PTC wavenumber band structure (Subsection \ref{subsec:bandstruct}) and an RWA for optical fields in time-varying media (Subsection \ref{subsec:rwa}).

\subsection{Photonic time crystal's wavenumber bandgap}
\label{subsec:bandstruct}

Because the PTC is spatially homogeneous, different wavenumbers $k$ remain uncoupled. Its temporal periodicity, however, gives Maxwell's equations Floquet solutions \cite{lyubarov2022amplified,asgari2024theory} that are periodic with the modulation period up to a factor $\exp[-i\omega_k^{(n)}t]$, where $\omega_k^{(n)}$ is a generally complex-valued Floquet frequency. The state space at each $k$ is two-dimensional, so there are only two indices $n$ for each wavenumber $k$. Furthermore, due to the time-reversal symmetry and the pseudo-Hermiticity of Maxwell's equations (Suppl.~Sec.~S5), (1) $\omega^{(1)}_k = -\omega^{(2)}_k \equiv \omega_k$; and (2) $\Re{\omega_k} = j\Omega/2$ for some integer $j$ inside the wavenumber bandgaps. Equivalently, $\exp(i2\pi \omega_k/\Omega)$ is either of magnitude one or purely real. If $\omega_k$ is purely real, $k$ is said to lie in a wavenumber (momentum) band. If $\omega_k$ is not purely real, $k$ is said to lie in a wavenumber (momentum) bandgap. Thus, wavenumber bandgaps appear only at $k$ such that $\omega_k = c_s |k| = j\Omega/2$ for integer $j$, as only then $\exp(i2\pi\omega_k/\Omega)$ is both purely real and of unit magnitude.

\begin{figure}
    \centering
    \includegraphics[width=\linewidth]{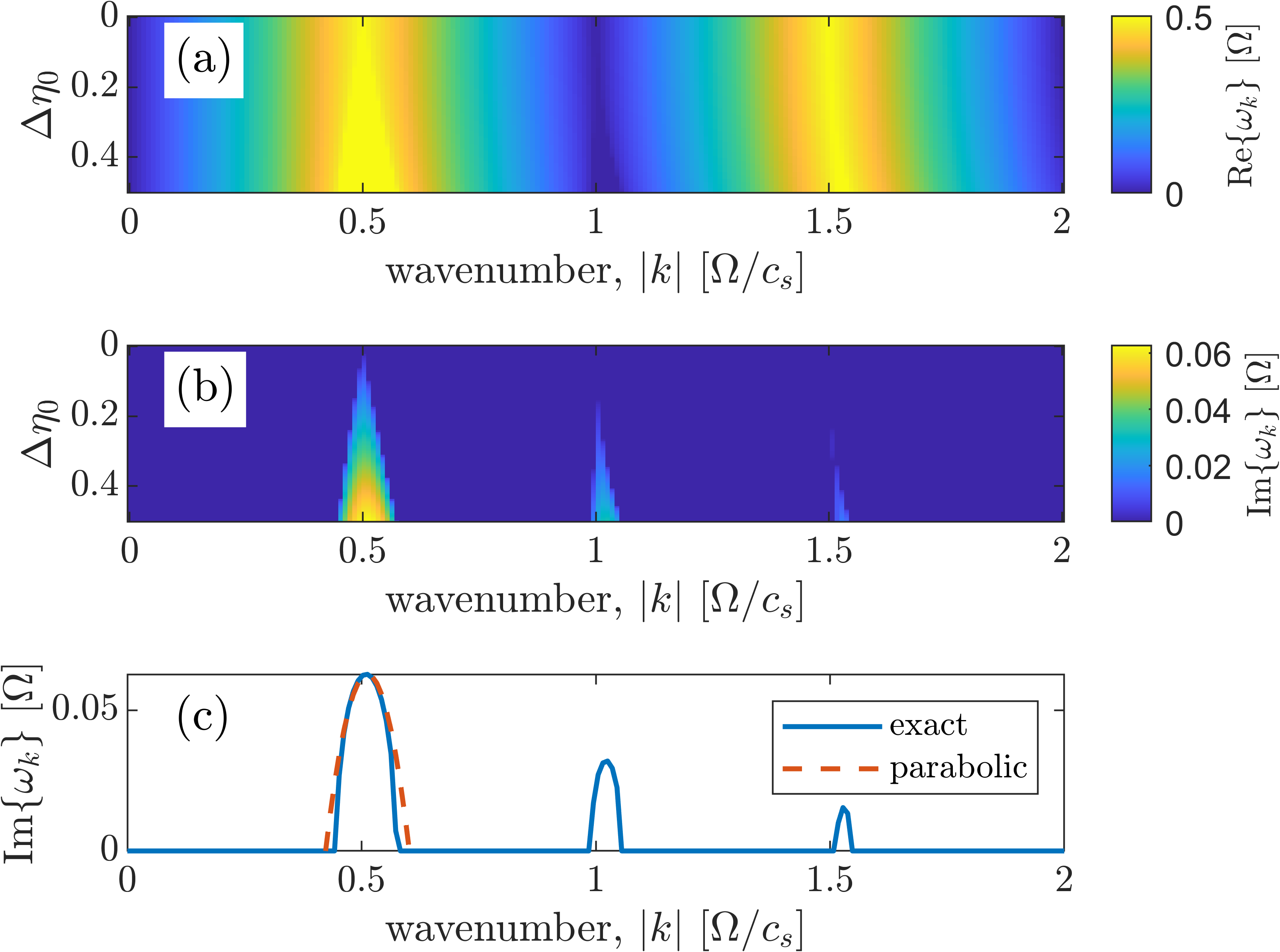}
    \caption{Floquet frequency $\omega_k$ as a function of wavenumber $k$ for variable and fixed modulation depth $\Delta\eta_0$. (a) Complex-valued Floquet frequency, $\omega_k$, for variable wavenumber $k$ and variable modulation, depth $\Delta\eta_0$. (b) Imaginary part of the Floquet frequency, $\Im{\omega_k}$, for variable wavenumber $k$ and fixed modulation depth, $\Delta\eta_0 = 1/2$}
    \label{fig:bandgap}
\end{figure}

Figures \ref{fig:bandgap}a and \ref{fig:bandgap}b show $\omega_k$ as a function of wavenumber $k$ and modulation depth $\Delta\eta_0$. As $\Delta\eta_0$ increases from zero, bandgaps with nonzero $\Im{\omega_k}$ emerge near $k=j\Omega/(2c_s)$, while $\Re{\omega_k}$ remains constant across each gap. Increasing modulation depth further broadens the gaps, increases their maximum $\Im{\omega_k}$, and shifts their peaks toward larger $k$.

Figure \ref{fig:bandgap}c shows $\Im{\omega_k}$ for $\Delta\eta_0 = 1/2$. Indexing each bandgap by $j=2c_sk_j/\Omega$, where $k_j$ is its weak-modulation peak, both the gap width and maximum $\Im{\omega_k}$ decrease with $j$. Moreover, $\Im{\omega_k}$ is smooth within each gap and is therefore locally parabolic near its maximum, a property important to interpret the results of Section \ref{sec:results}.

\subsection{Rotating-wave approximation for optical fields in time-varying media}
\label{subsec:rwa}

For each $k$, the PTC dynamics couple forward- and backward-propagating waves with unmodulated frequencies $\pm\omega_k=\pm c_s|k|$. This resembles a driven quantum-mechanical two-level system (QM-TLS) \cite{allen1975optical}: coupling accumulates resonantly when the modulation frequency matches the separation between the two natural frequencies (or, due to the additional frequency modulation induced by the PTC itself, any of its harmonics (Suppl.~Sec.~S6)). We therefore adapt the standard RWA of a driven QM-TLS to resonantly modulated optical fields in a PTC.

\begin{figure}
    \centering
    \includegraphics[width=\linewidth]{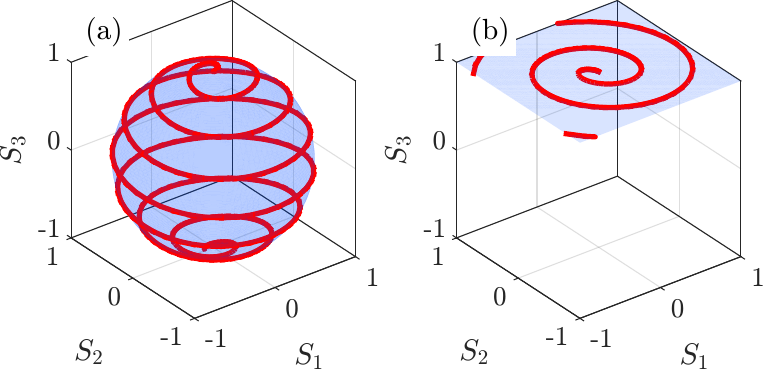}
    \caption{Geometrical representation of (a) Hermitian and (b) pseudo-Hermitian (with non-definite metric) time-evolutions in Bloch space}
    \label{fig:blochsphereplot2}
\end{figure}

The crucial distinction is that a QM-TLS is Hermitian, whereas PTC Maxwell dynamics are pseudo-Hermitian \cite{mostafazadeh2002pseudo,ashida2020nonhermitian,cortes2026pseudo}. A Hermitian system preserves a positive-definite metric, confining its Bloch trajectory to a compact surface (e.g.~a Bloch sphere). The PTC instead preserves a non-positive-definite Hermitian metric; its Bloch trajectory is confined only to a plane (Suppl.~Sec.~S5), a non-compact surface, and can therefore grow without bound, resulting in PTC amplification. Figure~\ref{fig:blochsphereplot2} illustrates these geometries, while the full PTC RWA derivation is given in Suppl.~Sec.~S6.

\section{Results and discussion}
\label{sec:results}

In this section, we examine the time-evolution of the figures of merit characterizing the spatial auto- and cross-correlations introduced in Section \ref{sec:spatiospectral}. We analyze their behavior by leveraging the analytical tools of Section \ref{sec:analytical}. We separate this section into two subsections. The first subsection (Subsection \ref{subsec:correlations}) analyzes the evolution of the fields' spatial correlations and spectra, alongside that of the figures of merit defined to describe them (i.e., their mean wavenumber, coherence length, and spatial bandwidth). The second subsection (Subsection \ref{subsec:matrices}) examines the evolution of the spatial correlation matrix and the cross-spectral density matrix.

For concreteness, we consider a PTC modulation with modulation depth $\Delta\eta_0$ of half the static impermittivity $\eta_s$ and a modulation frequency $\Omega$ of $2 k_B T/\hbar$ (see Sec.~\ref{sec:system} for their definition). Though the qualitative results we obtain, such as the convergence towards periodic asymptotics, are independent of these particular modulation parameters.

\subsection{Spatial correlations and spectral-densities}
\label{subsec:correlations}

\begin{figure*}
    \centering
    \includegraphics[width=\linewidth]{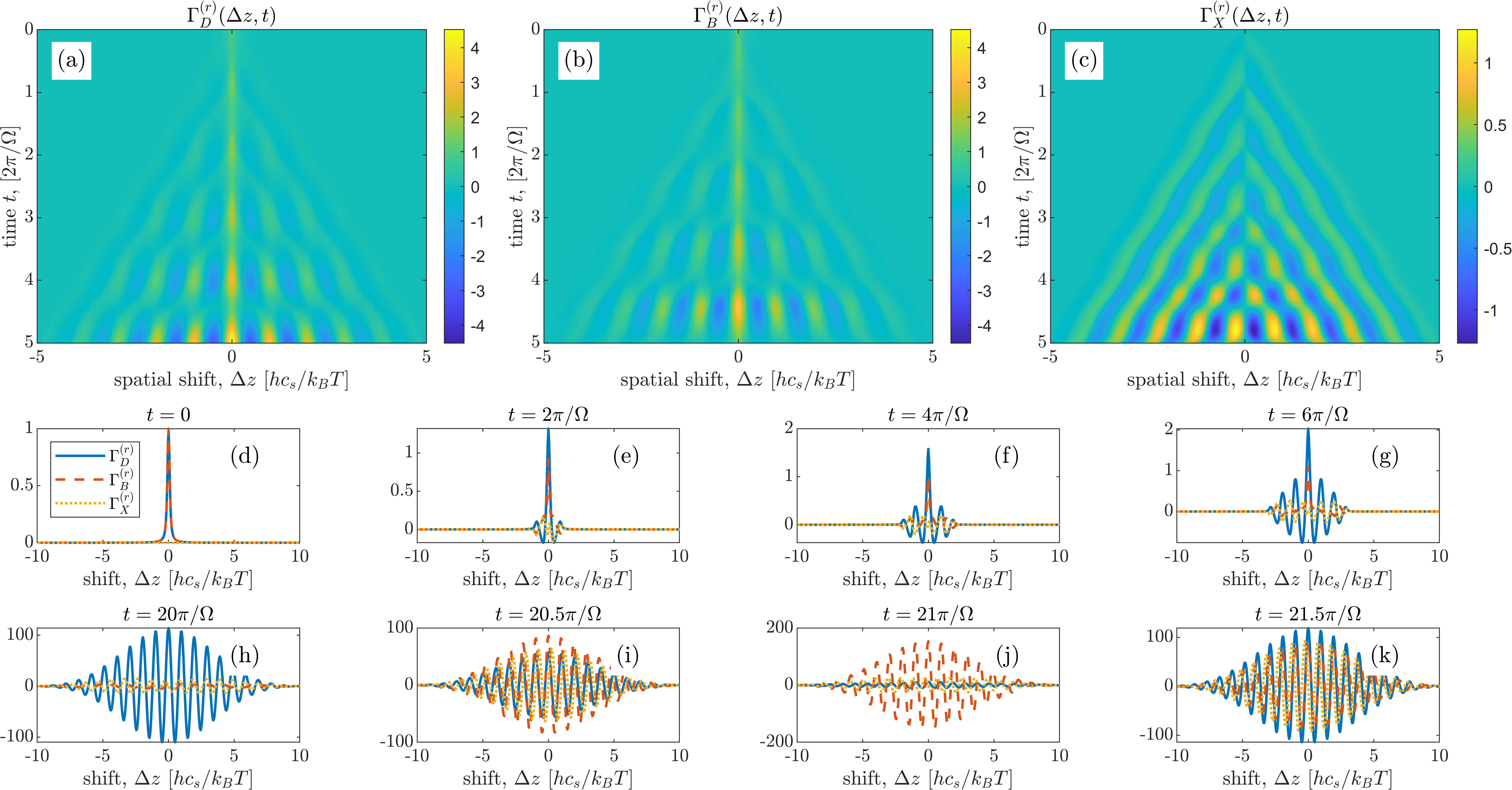}
    \caption{Evolution of the electromagnetic field's spatial correlations ($\Gamma^{(r)}_D$: displacement-field auto-correlation, $\Gamma^{(r)}_B$: magnetic-flux auto-correlation, $\Gamma^{(r)}_X$: displacement-field-magnetic-flux cross-correlation) due to PTC modulation. (a) $\Gamma_D$ for modulation time $t\in[0,10\pi/\Omega]$; (b) $\Gamma_B$ for $t \in [0, 10\pi/\Omega]$; and (c) $\Gamma_X$ for $t \in [0, 10\pi/\Omega]$. (d-k) field spatial correlations at different modulation times $t$.}
    \label{fig:ptcblackbodyfftdevalcontourfig2}
\end{figure*}

\begin{figure*}
    \centering
    \includegraphics[width=\linewidth]{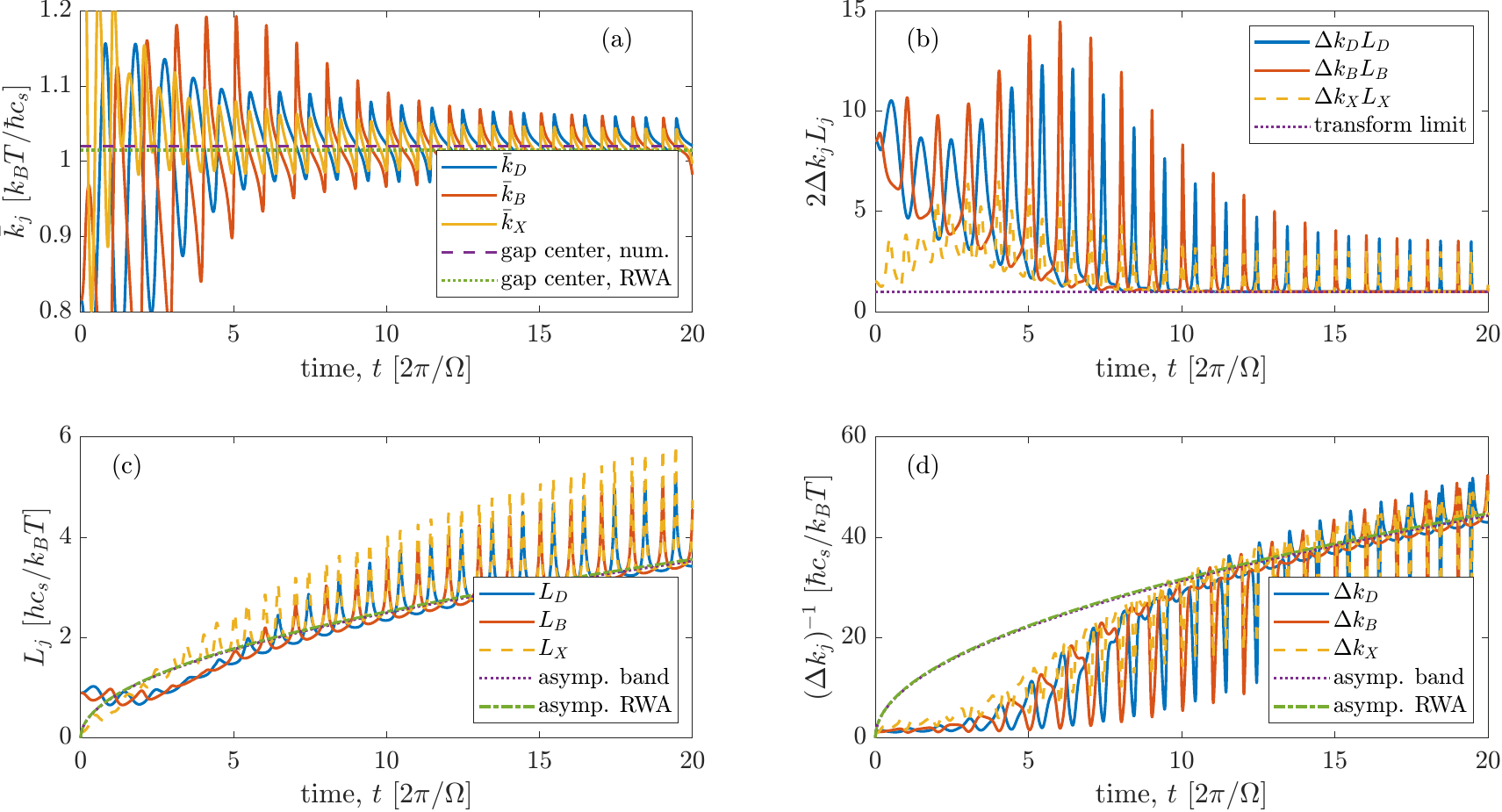}
    \caption{Non-trivial first and second moments of the fields' spatial correlations and spectral densities along with their long-time asymptotics: (a) mean wavenumbers, $\bar{k}_j$; (b) Coherence length-bandwidth products; $L_j \Delta k_j$; (c) coherence lengths, $L_j$; and (d) spatial bandwidths, $\Delta k_j$.}
    \label{fig:ptcblackbodyenergyrwacrossmat2fig1}
\end{figure*}

\begin{figure}
    \centering
    \includegraphics[width=\linewidth]{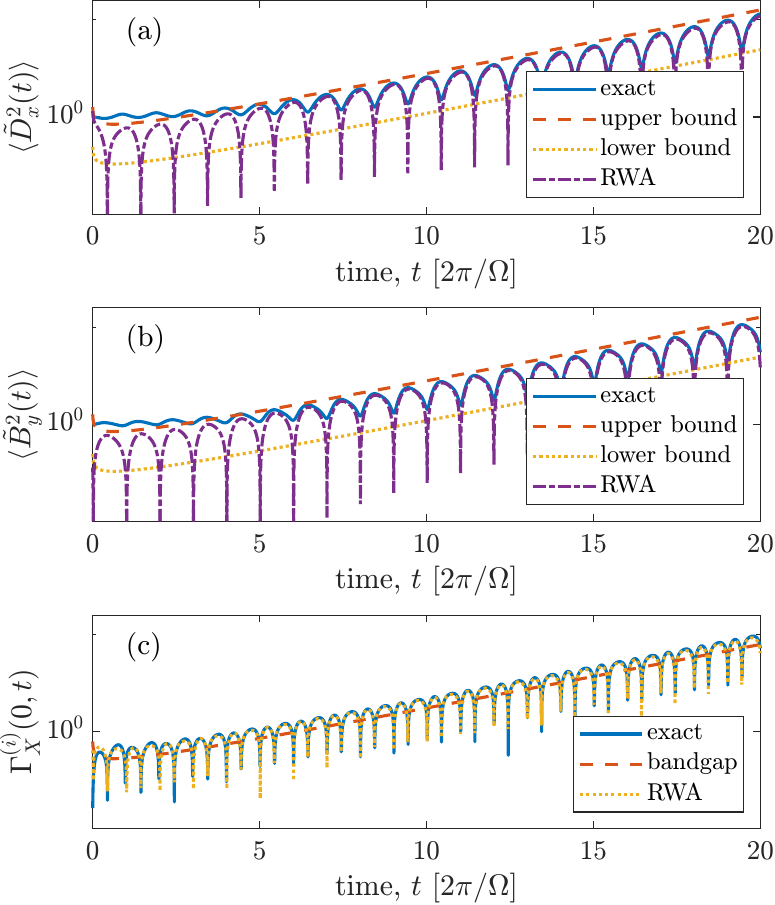}
    \caption{Field variances along with their long-time asymptotics obtained via the PTC's band-structure, and their estimates based on the rotating-wave approximation (RWA): (a) displacement field variance, (b) magnetix flux variance, and (c) displacement-field-magnetic-flux covariance (Hilbert transform).}
    \label{fig:ptcblackbodyenergyrwacrossmat2fig2}
\end{figure}

This subsection analyzes the evolution of the fields' spatial correlations $\Gamma_j(\Delta z,t)$ and spectra $W_j(k,t)$. Our objective is to understand how the field variances and cross-correlations and spatial correlations evolve over time, how the field variances are resolved over wavenumber, and how this evolution is governed by the PTC's band-structure and can be understood via the RWA. We do this analysis in two parts. First, we examine qualitatively the evolution of the spatial auto- and cross-correlations over time (Figure \ref{fig:ptcblackbodyfftdevalcontourfig2}). In this manner, we identify the main effects of PTC modulation on the spatial correlations, and we distinguish between two qualitatively distinct time regimes for their evolution: a transient regime, and an asymptotic regime. Second, we dissect quantitatively the evolution of the spatial correlations and spatial spectra through their (spectrally integrated) field variances (Figure \ref{fig:ptcblackbodyenergyrwacrossmat2fig2}) and through the figures of merit introduced in Section \ref{sec:spatiospectral}, i.e., their central wavenumbers, coherence lengths, and bandwidths (Figure \ref{fig:ptcblackbodyenergyrwacrossmat2fig1}). In this way, we show that, in the asymptotic (long-time) regime, the spatial correlations converge to a Gaussian shape with coherence length growing as square-root of the modulation time. We further identify periodic bursts when the spatial spectra diverge from this Gaussian shape, and we explain them in terms of the PTC band structure and the RWA.

Thus, we begin by analyzing the evolution of the real-valued spatial correlations $\Gamma^{(r)}_j(\Delta z,t)$, depicted in Figure \ref{fig:ptcblackbodyfftdevalcontourfig2}. These are computed numerically by evolving the double-sided spatial spectra $W_j(k,t)$ via Eq.~\eqref{eq:Wjtime}, and then Fourier transforming the result based on the Wiener-Khinchine theorem, Eq.~\eqref{eq:wienerkhinchine}. Initially, the spatial correlations are described by Eqs.~\eqref{eq:initial} and \eqref{eq:initial2}, with equal auto-correlations and vanishing cross-correlation. As the modulation time $t$ increases, the correlations oscillate and evolve gradually. In particular, the cross-correlation becomes non-vanishing, the field auto-correlations generally become non-degenerate (i.e., $\Gamma^{(r)}_D(\Delta z,t) \neq \Gamma^{(r)}_B(\Delta z,t)$), the field variances $\Gamma^{(r)}_D(0,t)$ and $\Gamma^{(r)}_B(0,t)$ increase, and all coherence lengths $L_j(t)$ increase.

The evolution of the correlations exhibits two qualitatively distinct regimes: a transient regime, and an asymptotic regime. For the PTC modulation of Fig.~\ref{fig:ptcblackbodyfftdevalcontourfig2}, the transition between these two occurs roughly at $t \sim 10 \pi /\Omega$. The evolution in the transient regime (Fig.~\ref{fig:ptcblackbodyfftdevalcontourfig2}(d) to \ref{fig:ptcblackbodyfftdevalcontourfig2}(g)) is irregular and hard to describe beyond the general observations made in the previous paragraph. In contrast, evolution in the asymptotic regime (Fig.~\ref{fig:ptcblackbodyfftdevalcontourfig2}h to \ref{fig:ptcblackbodyfftdevalcontourfig2}k) becomes more regular and easier to understand.

In the asymptotic regime, we observe that the correlations' oscillations become regular, almost sinusoidal in time with the frequency of the modulation and with well-defined phase. To see this, in Fig.~\ref{fig:ptcblackbodyfftdevalcontourfig2}h to \ref{fig:ptcblackbodyfftdevalcontourfig2}k, we plot the spatial correlations for $t = 20 \pi/\Omega$, $20.5\pi/\Omega$, $21\pi/\Omega$, $21.5\pi/\Omega$, i.e., every quarter of a period well in the asymptotic regime. We observe that $\expval{D^2_x(t)} = \Gamma^{(r)}_D(0,t)$ ($\expval{B^2_y(t)} = \Gamma^{(r)}_B(0,t)$) attains a local maximum (minimum) near $t = 20\pi/\Omega$, and and local minimum (maximum) at minimum near $21\pi/\Omega$. On the other hand, for fixed $\Delta z$, the cross-correlation $\Gamma^{(r)}_X(\Delta z,t)$ is attains a local magnitude-minimum near $t = 20\pi/\Omega$ and $t = 21\pi/\Omega$, and a local magnitude-maximum near $t = 20.5\pi/\Omega$ and $t = 21.5\pi/\Omega$, with these two magnitude maxima having different sign. Thus, in the asymptotic regime, $\Gamma^{(r)}_D(0,t)$ ($\Gamma^{(r)}_B(0,t)$) appears to oscillate as $(1 + \cos\Omega t)$ ($(1 - \cos\Omega t)$); and $\Gamma_X(\Delta z,t)$ for fixed, non-zero $\Delta z$, as $\sin\Omega t$. We further note from Fig.~\ref{fig:ptcblackbodyfftdevalcontourfig2}h to \ref{fig:ptcblackbodyfftdevalcontourfig2}k that, in the asymptotic regime, the spatial correlations increase almost exponentially with modulation time, and their envelopes with respect to spatial shift $\Delta z$ become approximately Gaussian. 

Next, we discuss the evolution of the spatial figures of merit introduced in Section \ref{sec:spatiospectral}: the mean wavenumbers $\bar{k}_j$, the coherence lengths $L_j$, and the spatial bandwidths $\Delta k_j$. The objective is to summarize the features of the spatial correlations $\Gamma_j(\Delta z,t)$ and spatial spectra $W_j(k,t)$ and thus enable a more precise description of their evolution due to PTC modulation. The time evolution of these figures of merit is presented in Figure \ref{fig:ptcblackbodyenergyrwacrossmat2fig1}.

We begin with the analysis of the evolution of the mean wavenumbers $\bar{k}_j(t)$. Those of the auto-correlations, $\bar{k}_D$ and $\bar{k}_B$, start at $0.8522 \times k_B T/ h c_s$ (Suppl.~Sec.~S4). That of the cross-correlation, $\bar{k}_X$ is ill-defined at $t = 0$. As observed for spatial correlations $\Gamma^{(r)}_j(\Delta z,t)$ in general, the mean wavenumbers exhibit a transient time regime, and an asymptotic time regime, separated again roughly at $t \sim 10 \pi /\Omega$. Overall, in the transient regime, the mean wavenumbers $\bar{k}_j$ oscillate with period $2\pi/\Omega$; although their evolution is irregular. In the asymptotic regime, the behavior becomes more regular, with the $\bar{k}_j$ converging to the main bandgap peak (Subsection \ref{subsec:bandstruct}), as shown in Fig.~\ref{fig:ptcblackbodyenergyrwacrossmat2fig1}a. This peak is accurately approximated by the RWA, with only a slight underestimation, making the RWA expression a simpler and effective alternative for analysis (Suppl.~Sec.~S7).

Owing to its simpler behavior, we focus on the asymptotic regime of the mean wavenumbers $\bar{k}_j$, starting with $\bar{k}_D$. This mean wavenumber reaches local maxima at half-integer values of $\Omega t/(2\pi)$, where it rises sharply from a local minimum. Between these sharp transitions, $\bar{k}_D$ decreases smoothly and monotonically with time, and exhibits an inflection point at each integer value of $\Omega t/(2\pi)$. At these inflection points, $\bar{k}_D$ coincides with the main bandgap peak and lies midway between the preceding maximum and the following minimum. As the modulation time increases, the separation between successive extrema and the corresponding inflection point decreases monotonically.

The asymptotic behavior of $\bar{k}_D(t)$ can be understood from the evolution of $\Gamma^{(r)}_D(\Delta z,t)$ in Fig.~\ref{fig:ptcblackbodyfftdevalcontourfig2}h to \ref{fig:ptcblackbodyfftdevalcontourfig2}k. The sharp increments in $\bar{k}_D(t)$ at half-integer $\Omega t/(2\pi)$ coincide with local minima of the displacement variance, $\expval{D^2_x(t)} = \Gamma_D(0,t)$. In contrast, the inflection points at integer values coincide with variance maxima. These results indicate that the abrupt increases in $\bar{k}_D(t)$ arise from short-lived periodic transformations of $\Gamma_D^{(r)}(\Delta z ,t)$. While the variance reaches its maximum when the one-sided spectrum $W_D(k,t)$ is centered at the main bandgap peak.

Next, we discuss the asymptotic behavior of $\bar{k}_B$ and $\bar{k}_X$. The magnetic mean wavenumber $\bar{k}_B$ follows the same pattern as  $\bar{k}_D$, but with a phase shift of $\pi$, as shown in Figs. \ref{fig:ptcblackbodyfftdevalcontourfig2}h-\ref{fig:ptcblackbodyfftdevalcontourfig2}k and \ref{fig:ptcblackbodyenergyrwacrossmat2fig1}a; i.e., its sharp increases occur at integer values of $\Omega t/(2\pi)$, while its inflection points occur at half-integer values. In contrast, $\bar{k}_X$ exhibits sharp increases at both integer and half-integer values, doubling the oscillation frequency. Between successive increases, however, $\bar{k}_X$ displays the same monotonic decay and inflection-point behavior as $\bar{k}_D$ and $\bar{k}_B$.

As we discuss below for Figure \ref{fig:ptcblackbodyenergyrwacrossmat2fig2} and in Subsection \ref{subsec:matrices}, the sharp increases in $\bar{k}_j$ ($j = D,B,X$) occur when $|W_j(k,t)|$ near the main bandgap peak reaches a local minimum in time $t$. At these times, the integral expression for $\bar{k}_j$ is susceptible to contributions from the bands and other bandgaps. 


For the uncertainty products $L_j(t)\Delta k_j(t)$ (($j=D,B,X$); Fig.~\ref{fig:ptcblackbodyenergyrwacrossmat2fig1}b), PTC modulation induces periodic oscillations in spatial spectral complexity at the modulation frequency. Their overall evolution differs markedly between the transient and asymptotic regimes, separated at approximately $t \approx 10\pi/\Omega$. As in Fig.~\ref{fig:ptcblackbodyenergyrwacrossmat2fig1}a, the transient behavior is difficult to characterize because the figures of merit remain strongly influenced by the initial field conditions, while the amplified bandgap modes have not yet dominate and thus the non-amplified modes outside the main bandgap still contribute appreciably.

In contrast, the asymptotic regime exhibits a more regular evolution of the complexities $L_j(t)\Delta k_j(t)$. After reaching sharp global maxima at $t \approx 10\pi/\Omega$, the complexities oscillate periodically, with narrow peaks and broader valleys. The peaks coincide with rapid increases in the mean wavenumber $\bar{k}_j$. Both peak and valley amplitudes gradually decrease, converging to 2 and the Fourier transform limit of 1/2, respectively. Since this limit is attained only for Gaussian spectra (or equivalently, Gaussian spatial correlations), the spectra become approximately Gaussian at sufficiently long times ($t \gtrsim 20\pi/\Omega$), except during the narrow complexity peaks.

We attribute the asymptotic convergence of the spatial spectra to Gaussian profiles to the Bloch parameters, $S_j(k,t)$. At long modulation times, the $S_j(k,t)$ become approximately Gaussian with a spatial bandwidth proportional to $1/\sqrt{t}$ (Suppl.~Sec.~S7), causing $W_j(k,t)$ to inherit the same Gaussian form. The complexity peaks coincide with the sharp increases in the mean wavenumber $\bar{k}_j$, indicating a common origin in the transient reshaping of the spectra. At these times, $W_j(k,t)$ is strongly suppressed near the main bandgap center, and the spectra become dominated by the passbands and higher-order bandgaps.


Next, we examine the coherence lengths $L_j(t)$ ($j=D,B,X$; Fig.~\ref{fig:ptcblackbodyenergyrwacrossmat2fig1}c. In contrast to the complexity measures, the coherence lengths reach their asymptotic behavior more rapidly, around $t \approx 10\pi/\Omega$. In the asymptotic regime, $L_j$ oscillate periodically while their overall magnitudes gradually increase with time. Narrow maxima occur when $W_j(k,t)$ is minimized (at half-integer values of $\Omega t/2\pi$ for $L_D$, integer values for $L_B$, and both for $L_X$), separated by broader minima. Both maxima and minima increase approximately as $\sqrt{t}$. The cross-correlation coherence length $L_X$ is consistently about 10\% larger than $L_D$ and $L_B$. While the minima of all three coherence lengths follow a common envelope, the maxima of $L_D$ and $L_B$ share one envelope that differs from that of $L_X$.

We estimate the envelope of the coherence-length minima by noting that, at long modulation times, the spatial spectra become approximately Gaussian with a bandwidth proportional to $1/\sqrt{t}$ (Suppl.~Sec.~S7). Consequently, the spatial correlations are also Gaussian, with coherence lengths scaling as $\sqrt{t}$. The corresponding proportionality constant is obtained both numerically from the bandgap structure and semi-analytically utilizing the RWA (Suppl.~Sec.~S7). As shown in Fig.~\ref{fig:ptcblackbodyenergyrwacrossmat2fig1}c, the two approaches agree closely and accurately capture the asymptotic evolution of the coherence-length minima.


Next, we examine the spatial bandwidths $\Delta k_j$, plotted as $(\Delta k_j)^{-1}$ in Fig.~\ref{fig:ptcblackbodyenergyrwacrossmat2fig1}d. Their transient regime extends to $t \approx 20\pi/\Omega$, matching that of the uncertainty products $L_j\Delta k_j$, indicating that the latters' transient regime are governed primarily by the bandwidths rather than the coherence lengths. In the asymptotic regime, $(\Delta k_j)^{-1}$ exhibits periodic sharp dips separated by broad valleys. The dips occur when $W_j(k,t)$ are minimized near the main bandgap center. Successive dips in $(\Delta k_D)^{-1}$ and $(\Delta k_B)^{-1}$ are comparable and increase gradually with time, whereas those of $(\Delta k_X)^{-1}$ remain consistently smaller. Superimposed on these periodic oscillations, the spatial bandwidths decrease gradually over time.

As for the coherence lengths, the asymptotic Gaussian form of the spatial spectra enables simple estimates of the spatial bandwidths $\Delta k_j$ (Suppl.~Sec.~S7). The bandwidths are found to scale as $1/\sqrt{t}$, with the proportionality constant obtained both numerically from the PTC bandgap structure and semi-analytically via the RWA. As shown in Fig.~\ref{fig:ptcblackbodyenergyrwacrossmat2fig1}d, the two approaches agree closely and accurately reproduce the upper envelopes of $(\Delta k_j)^{-1}$, where the spatial spectra are nearly Gaussian.


Next, we examine the field variances, $\langle D^2(t)\rangle=\Gamma_D^{(r)}(0,t)$ and $\langle B^2(t)\rangle=\Gamma_B^{(r)}(0,t)$, which, by the Wiener–Khinchin theorem, equal the spectrally integrated power spectral densities. For the cross-spectral density $W_X(k,t)$, the corresponding integrated quantity is $\langle D_x(t)B_y(t)\rangle$, the ensemble-averaged Minkowski momentum. However, this quantity remains zero because spatially homogeneous PTC modulation conserves Minkowski momentum \cite{ortega2023tutorial}, and the initial momentum vanishes by equipartition \cite{haus2000electromagnetic}. Therefore, although $\langle D_x(z,t)B_y(z+\Delta z,t)\rangle$ is generally nonzero for $\Delta z\neq0$ (Fig. 5), we instead analyze $\Gamma_X^{(i)}(0,t)$, the Hilbert transform of $\Gamma_X^{(r)}(\Delta z,t)$ evaluated at $\Delta z=0$, which provides an integrated measure of the cross-field correlation from the definition of the Hilbert transform \cite{bracewell2000fourier,mandel1995optical,wolf2007introduction}.

We thus examine the effective field variances $\expval{\tilde{D}^2(t)}$, $\expval{\tilde{B}^2(t)}$, and $\Gamma_X^{(i)}(0,t)$ (Fig.~\ref{fig:ptcblackbodyenergyrwacrossmat2fig2}). After a short transient, all three quantities oscillate with the modulation period while their amplitudes grow approximately as
\begin{equation}
    \exp(2\alpha_0 t) /\sqrt{t}, \label{eq:variancetdepend}
\end{equation}
where $\alpha_0$ is the Floquet amplification rate at the main bandgap peak (Suppl.~Sec.~S7). The factor of $1/\sqrt{t}$ reflects that only the exact bandgap peak is amplified at the maximum rate $\alpha_0$. The local minima of the effective variances coincide with the minima of the corresponding spatial spectra $W_j(k,t)$ near the main bandgap, analogous to the sharp features identified in Fig.~\ref{fig:ptcblackbodyenergyrwacrossmat2fig1}.

To further dissect the effective field variances, we develop two approximation schemes (Suppl.~Sec.~S7). The first is based on an eigenvalue decomposition of the one-period evolution matrix $U(k,2\pi/\Omega)$ [Eq.~\eqref{eq:Ukt}], retaining only the dominant Floquet eigenmode. This approach yields tight upper and lower bounds for $\tilde{D}^2(t)$ and $\tilde{B}^2(t)$, and the modulation-period-averaged evolution of $\Gamma_X^{(i)}(0,t)$. It confirms the asymptotic scaling of Eq.~\eqref{eq:variancetdepend} and determines the corresponding time-independent prefactors for each effective variance.


The second scheme applies the RWA (Subsection \ref{subsec:rwa}). We derive approximate analytical expressions for the Bloch components $S_j(k,t)$, expand the resulting spectra $W_j(k,t)$ to second order around the main bandgap peak, and integrate over wavenumber (Suppl.~Sec.~S7). The resulting RWA curves in Fig.~\ref{fig:ptcblackbodyenergyrwacrossmat2fig2} accurately capture the oscillations and semi-exponential growth of the variances. Their main discrepancy occurs near the variance minima, where contributions from wavenumbers outside the main bandgap become important and the RWA is less accurate.

\subsection{Spatial-correlation matrix and cross-spectral matrix}
\label{subsec:matrices}

In this section, we extend the analysis of Subsection \ref{subsec:correlations} to arbitrary linear combinations of the displacement field $D_x(z,t)$ and magnetic flux $B_y(z,t)$ by examining the spatial cross-correlation matrix $\bar{\Gamma}(\Delta z,t)$ and cross-spectral matrix $\bar{W}(k,t)$ (Subsection \ref{subsec:definitions}).

\begin{figure*}
    \centering
    \includegraphics[width=\linewidth]{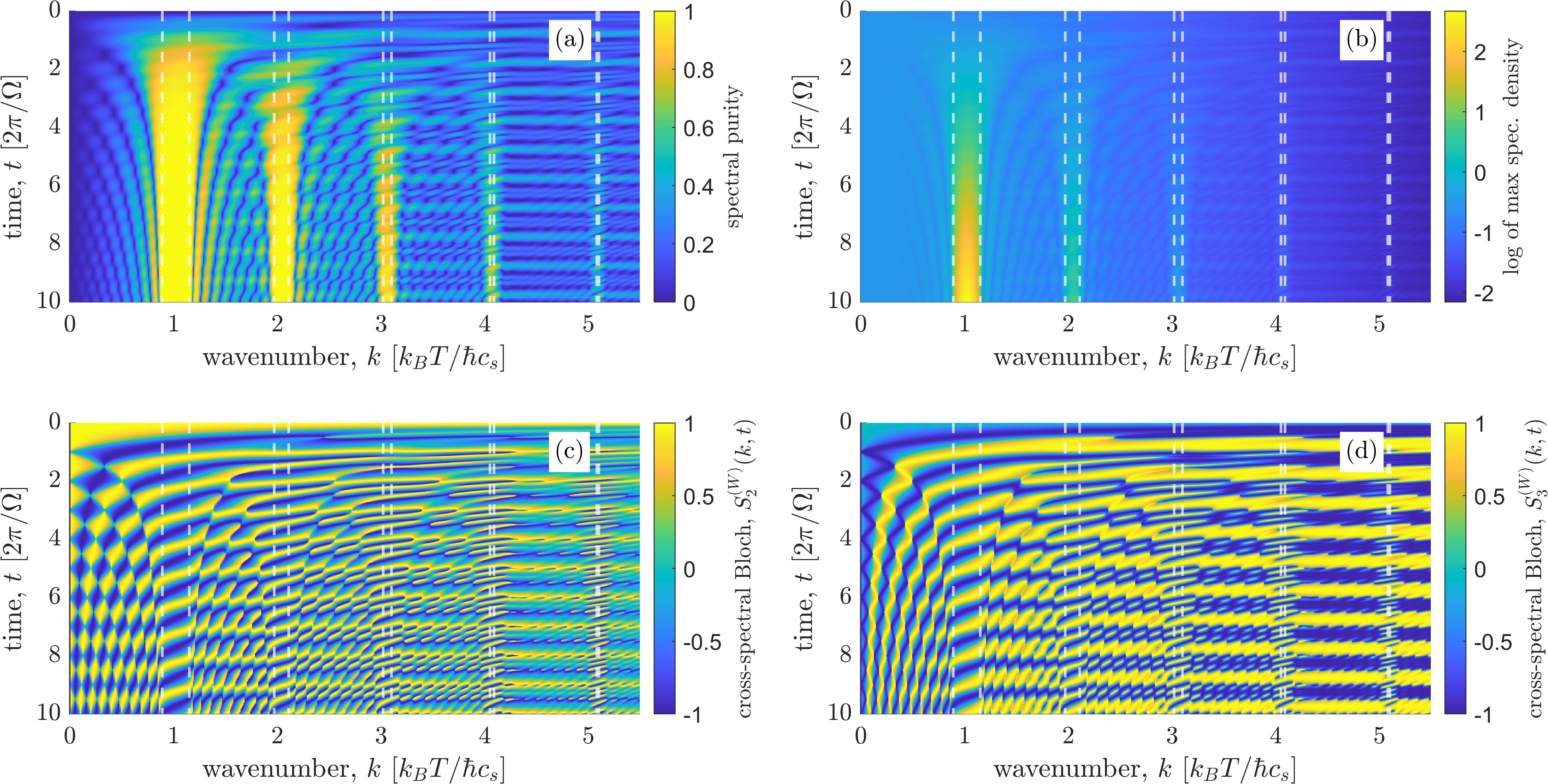}
    \caption{Eigenvalue decomposition of the cross-spectral density matrix $\bar{W}$ as a function of wavenumber $k$ and modulation time $t$: (a) spectral purity; (b) maximum spectral density; and (c-d) cross-spectral density Bloch vector components,  $S^{(W)}_2$ (c) and $S^{(W)}_3$ (d). Edges of the PTC's bandgaps are shown as white dashed lines.}
    \label{fig:ptcblackbodyenergyrwacrossmat2}
\end{figure*}

We first analyze $\bar{W}(k,t)$ through its eigenvalue decomposition (Fig.~\ref{fig:ptcblackbodyenergyrwacrossmat2}). As before, its evolution exhibits two distinct regimes: a transient regime and an asymptotic regime, separated at approximately $t\approx10\pi/\Omega$. In the asymptotic regime, $\bar{W}(k,t)$ evolves differently within the PTC bands and bandgaps. During the transient regime, apparent amplification and deamplification occur near, but outside, the bandgaps (delimited by white dashed lines in Fig.~\ref{fig:ptcblackbodyenergyrwacrossmat2}). These effects vanish at long times, as detuning between the modulation frequency and the natural mode frequencies suppresses off-resonant amplification.

We now focus on the asymptotic regime. Within the PTC bandgaps, both the spatial purity and the maximum spectral density increase monotonically, with slower growth for higher-order bandgaps (larger central wavenumbers). In contrast, inside the PTC's passbands these quantities do not accumulate but instead oscillate at both the modulation frequency and the mode's natural frequency, $c_s |k|$. These oscillations indicate weak coherent amplification that is suppressed by the detuning between the modulation and natural frequencies outside the momentum bandgaps.


Next, we examine the eigenvectors of the cross-spectral matrix $\bar{W}(k,t)$. As explained in Subsection \ref{subsec:definitions}, it suffices to consider the Bloch-vector components $S^{(W)}_j(k,t)$  of the eigenvector with largest eigenvalue. Because the cross-spectrum $W_X(k,t)$ is purely imaginary, consistent with conservation of Minkowski momentum \cite{ortega2023tutorial}, $S_2^{(W)}(k,t)$ vanishes, leaving only $S_1^{(W)}$ and $S_3^{(W)}$. As shown in Figs.~\ref{fig:ptcblackbodyenergyrwacrossmat2}c and \ref{fig:ptcblackbodyenergyrwacrossmat2}d, their evolution differs markedly between the bandgaps and passbands (separated by white dashed lines in Fig.~\ref{fig:ptcblackbodyenergyrwacrossmat2}). Within each bandgap, the Bloch components oscillate at a uniform frequency with a smoothly varying phase and an inflection point near the bandgap center; the oscillation frequency increases with bandgap order. In the passbands, the Bloch components instead fluctuate at both the modulation frequency and the natural mode frequency, reflecting the suppression of coherent amplification by off-resonant detuning.

\begin{figure*}
    \centering
    \includegraphics[width=\linewidth]{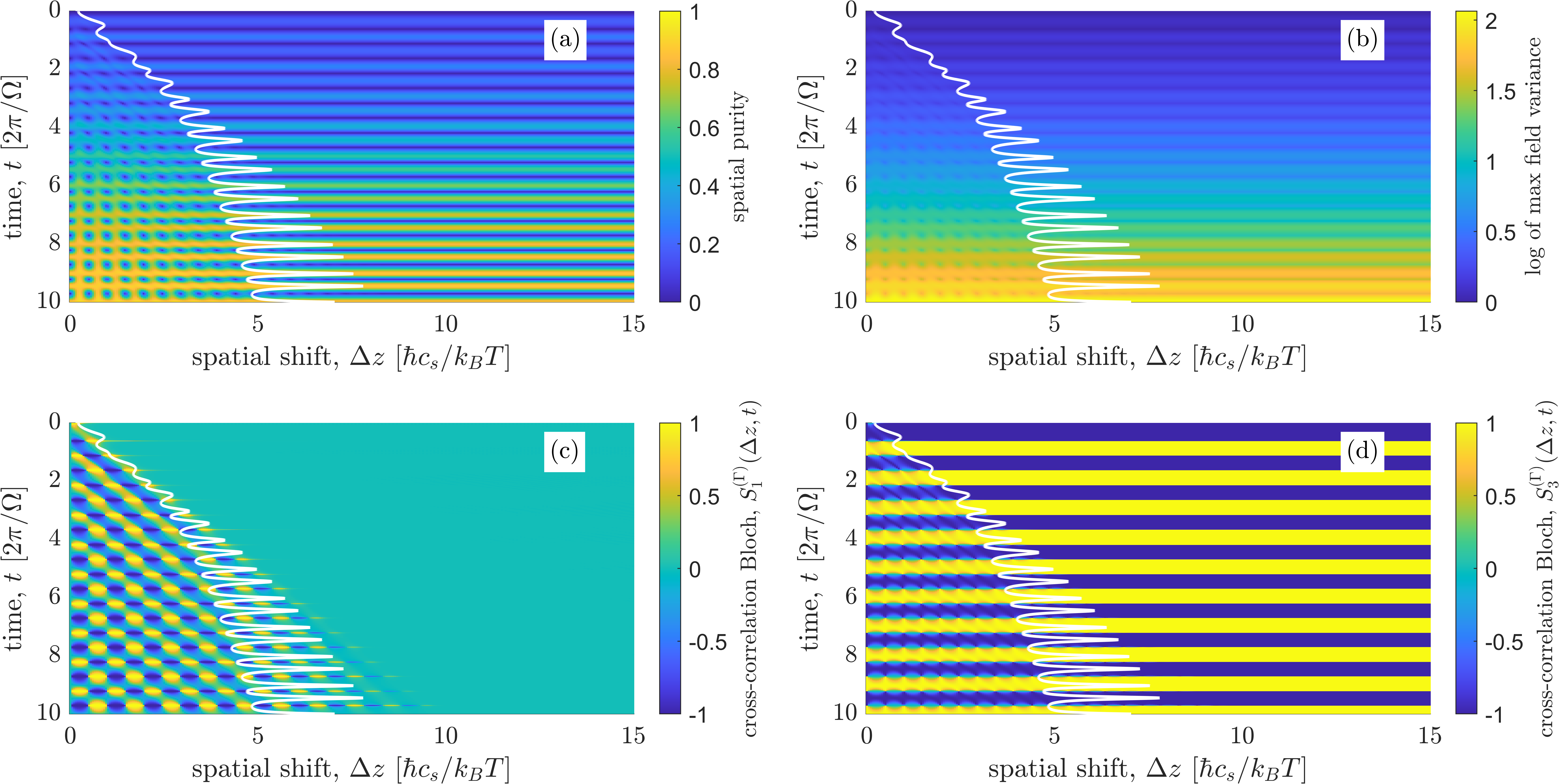}
    \caption{Eigenvalue decomposition of the spatial cross-correlation matrix $\bar{\Gamma}$ as a function of spatial shift $\Delta z$ and modulation time $t$: (a) spatial purity; (b) maximum field variance; and (c-d) Spatial cross-correlation Bloch-vector components, $S^{(\Gamma)}_1$ (c), and $S^{(\Gamma)}_3$ (d). White solid lines correspond to $2 L_X(t)$.}
    \label{fig:ptcblackbodyenergyrwacrossmat214}
\end{figure*}

We next examine the spatial correlation matrix $\bar{\Gamma}(\Delta z,t)$ (Subsection \ref{subsec:definitions}), characterized by its spatial purity, maximum field variance, and Bloch-vector components (Fig.~\ref{fig:ptcblackbodyenergyrwacrossmat214}). These quantities exhibit two qualitatively distinct regions in the $(\Delta z,t)$-plane, roughly delimited by twice the cross-coherence length, $2L_X(t)$ (white continuous line in Fig.~\ref{fig:ptcblackbodyenergyrwacrossmat214}). In this $(\Delta z,t)$-plane, we thus define a Region 1 by $|\Delta z|\le 2L_X(t)$ and a Region 2 by $|\Delta z|> 2L_X(t)$. We note that this separation is just rough because of two reasons. First, because $\bar{\Gamma}(\Delta z,t)$ depends on $\Delta z$ through the cross-correlation $\Gamma^{(r)}(\Delta z,t)$, which varies continuously and smoothly with $\Delta z$ rather than discretely. And second, because the shape of the cross-correlation $\Gamma^{(r)}(\Delta z,t)$ as a function of $\Delta z$ changes over the transient regime (roughly, $t \leq 10\pi /\Omega$), so the decay of $\Gamma^{(r)}_X(\Delta z,t)$ over a coherence length $L_X(t)$ changes gradually over this transient regime.

In Region 2, the cross-correlation $\Gamma^{(r)}_X(\Delta z,t)$ is negligible, so $\bar{\Gamma}(\Delta z,t)$ is dominated by the dielectric and magnetic field variances, $\Gamma^{(r)}_D(0,t)$ and $\Gamma^{(r)}_B(0,t)$, and is therefore nearly independent of $\Delta z$, varying only with modulation time $t$. In Region 1, however, the cross-correlations become non-negligible and make $\bar{\Gamma}$ strongly dependent on $\Delta z$. An exception occurs when either $\langle D_x^2\rangle$ or $\langle B_y^2\rangle$ reaches a maximum, as the remaining variance and cross-correlation are then minimal, rendering $\bar{\Gamma}$ nearly independent of $\Delta z$. At times when the two field variances are comparable, the nonzero cross-correlation $\Gamma_X(\Delta z,t)$ lifts the eigenvalue degeneracy of $\bar{\Gamma}$, causing its properties to oscillate with $\Delta z$. Because $\Gamma_X(\Delta z,t)$ asymptotically approaches a Gaussian envelope modulated at the bandgap wavenumber (Subsection \ref{subsec:correlations}), the same spatial envelope and oscillation period emerge in the properties of $\bar{\Gamma}$.


We first examine the spatial purity $P_\Gamma(\Delta z,t)$. Initially, $P_\Gamma(\Delta z,0)=0$ because the electric and magnetic fields have equal energies and are uncorrelated, consistent with equipartition. As the PTC modulation proceeds, the field variances become unequal and the cross-correlation develops, causing $P_\Gamma$ to increase with $t$.

In Region 2 ($|\Delta z|>L_X(t)$), the cross-correlation is negligible, so $P_\Gamma$ depends only on the imbalance between the electric and magnetic variances. Consequently, it is approximately independent of $\Delta z$, oscillates with the modulation frequency, and its maxima approach the limiting value $P_\Gamma=1$. In Region 1 ($|\Delta z|\le L_X(t)$), the non-vanishing cross-correlation makes $P_\Gamma$ spatially dependent. When the field variances are nearly equal, $\Gamma_X(\Delta z,t)$ lifts the eigenvalue degeneracy of $\bar{\Gamma}$, producing oscillations of $P_\Gamma$ with $\Delta z$. These oscillations follow the Gaussian envelope and spatial period of $\Gamma_X(\Delta z,t)$.

To characterize the eigenvalue scale, we also consider the largest eigenvalue, $\gamma_+$ (Fig.~\ref{fig:ptcblackbodyenergyrwacrossmat214}b). It depends primarily on time, oscillating with the modulation period while increasing nearly exponentially, consistent with the field variances in Fig.~\ref{fig:ptcblackbodyenergyrwacrossmat2fig2}. Near times when the electric and magnetic variances coincide, the finite cross-correlation in Region 1 lifts the eigenvalue degeneracy, leading to a local increase in $\gamma_+$.



The corresponding eigenvectors are described by the Bloch-vector components  $S^{\Gamma}_j(\Delta z,t)$ (Subsection \ref{subsec:definitions}). Because $\bar{\Gamma}$ is real-valued, $S_2^{(\Gamma)}=0$. In Region 2, where $\Gamma_X\approx 0$, then $S_1^{(\Gamma)}=0$ and $S_3^{(\Gamma)}=\pm 1$, depending on which field variance dominates. In Region 1, the non-vanishing cross-correlation makes $S_1^{(\Gamma)}$ oscillate with $\Delta z$, while $|S_3^{(\Gamma)}|$ decreases to preserve the unit Bloch-vector magnitude (Subsection \ref{subsec:definitions}).



These results provide a physical picture of PTC amplification of blackbody radiation. The initially thermal field has equal electric and magnetic energies and zero cross-correlation. PTC modulation breaks this balance, generating standing-wave correlations while conserving the initial zero Minkowski momentum. Constructive interference within the PTC bandgaps amplifies these correlations, increasing both the spatial coherence and spatial purity of the field.



\section{Conclusion}
\label{sec:conclusion}

We investigated the evolution of initially thermal (blackbody) radiation in a one-dimensional photonic time crystal (PTC). Starting from Maxwell's equations in spatially homogeneous, time-modulated media, we derived the evolution of the fields' spatial correlations and power spectral densities. Their dynamics were interpreted using the PTC band structure and a rotating-wave approximation describing the pseudo-Hermitian dynamics of nearly resonant radiation. We showed that the field spectra exhibit a universal asymptotic behavior governed by the main PTC bandgap: the initially thermal radiation periodically evolves toward Gaussian spatial spectra and correlations, accompanied by increasing amplitude, coherence length, and spatial- and wavenumber-domain purity. Beyond providing fundamental insight into PTC amplification of a stochastic electromagnetic field, this work offers guidance for understanding and engineering the amplification of thermal radiation without or alongside coherent electromagnetic radiation.


\section*{Acknowledgements}
We thank Prof.~Miguel A.~Alonso for fruitful discussions about this work and its analogy to the theory of partial polarization. L.~C., J.~H, and Y.~X. acknowledge support from DARPA YFA program (Grant No. D23AP0018900) and NSF LEAPS-MPS program (Award Number 2418002). N.~G. acknowledges support from the Undergraduate Research Fellowship from University of North Texas.


\bibliographystyle{ieeetr}
\bibliography{thermalptcbib}

\clearpage


\section*{Supplementary material}

# Temporal Evolution of Blackbody Radiation in a One-Dimensional Photonic Time Crystal: Supplemental Document

## S1. DISCUSSION OF THE CLASSICAL NATURE OF BLACKBODY RADIATION IN PHOTONIC TIME CRYSTAL

In this section, we discuss sufficient conditions under which initially blackbody radiation in a photonic time crystal (PTC) can be described as a classical random process and not just quantum-mechanically. Quantum mechanically, blackbody radiation consists of a quasi-continuum of electromagnetic modes each in their thermal states at some temperature $T$ [S1]. Thermal states belong to the family of so-called Gaussian states [S2, S 3], i.e., quantum-optical states which Wigner functions are Gaussian functions. It is known that all Gaussian states remain Gaussian under time evolution induced by any bilinear Hamiltonian [S3]. Furthermore, the Hamiltonian of light in a PTC is indeed bilinear [S4]. Hence all modes initially in thermal states evolve remaining Gaussian states.

Gaussian states are fully determined by their first-moments and covariances, and these first-moments and covariances evolve as classical phase distributions [S3]. Thus, if the initial first-moments and covariances correspond to classical states, they will continue to do so after time evolution in a PTC. For a thermal state, the first moments vanish, and the covariance matrix is diagonal with equal variances $(N_k + 1/2)$ [S3] ($N_k$ is the mean number of thermal photons). This variance corresponds to a classical state when $N_k \gg 1$, i.e., it is practically unconstrained by the uncertainty relations between canonical operators. Thus, modes with initial states of large $N_k$ remain well described by classical fields in PTCs.

Of course, for sufficiently large wavenumber $|k|$, eventually $\hbar\omega_k = \hbar c_s |k| \sim k_B T$, i.e., the photon energy becomes comparable to the thermal energy; so the initial state is quantum-mechanical. However, we assume the PTC's modulation frequency $\Omega$ is such that these high-wavenumber modes accumulate no amplification, i.e., they undergo merely adiabatic evolution [S5]. Hence, even under a fully quantum-mechanical theory, their contribution to spatial correlations and spectrally integrated figures of merit (such as those of the manuscript) is negligible.

## S2. REPRESENTATION OF 1D BLACKBODY RADIATION AS A CLASSICAL RANDOM PROCESS

In this section, we discuss how to represent blackbody radiation as a classical random process to the accuracy of Section S1, while retaining its bandwidth and field correlations.

Under full field quantization, an optical field's electric $\hat{\mathbf{E}}(\mathbf{x}, t)$ and magnetic $\hat{\mathbf{B}}(\mathbf{x}, t)$ in a static and spatially homogeneous dielectric medium are given in the Heisenberg picture by [S1, S 6]

$$\begin{aligned} \hat{\mathbf{E}}(\mathbf{x}, t) &= \hat{\mathbf{E}}^{(+)}(\mathbf{x}, t) + \text{H. c.}, \\ \hat{\mathbf{B}}(\mathbf{x}, t) &= \hat{\mathbf{B}}^{(+)}(\mathbf{x}, t) + \text{H. c.} \end{aligned} \tag{S1}$$

Here, $\hat{\mathbf{E}}^{(+)}(\mathbf{x}, t)$ and $\hat{\mathbf{B}}^{(+)}(\mathbf{x}, t)$ are the so-called positive-frequency components of the electric and magnetic field, respectively, given by [S1, S 6]

$$\begin{aligned} \hat{\mathbf{E}}^{(+)}(\mathbf{x}, t) &= \frac{i}{\sqrt{AL_z}} \sum_{\mathbf{k},s} \sqrt{\frac{\hbar\omega_{\mathbf{k}}}{2\varepsilon_s}} \hat{a}_{\mathbf{k},s} \boldsymbol{\varepsilon}_{\mathbf{k},s} \exp\left(i\mathbf{k}\cdot\mathbf{x} - i\omega_{\mathbf{k}} t\right), \\ \hat{\mathbf{B}}^{(+)}(\mathbf{x}, t) &= \frac{i}{c_s\sqrt{AL_z}} \sum_{\mathbf{k},s} \sqrt{\frac{\hbar\omega_{\mathbf{k}}}{2\varepsilon_s}} \hat{a}_{\mathbf{k},s} (\hat{\mathbf{k}} \times \boldsymbol{\varepsilon}_{\mathbf{k},s}) \exp\left(i\mathbf{k}\cdot\mathbf{x} - i\omega_{\mathbf{k}} t\right). \end{aligned} \tag{S2}$$

Here, $\hbar$ is Planck's reduced constant; $L_z$, the length of the quantization volume along its $z$ axis (longitudinal coordinate in the manuscript and following sections); $A$, the quantization volume's cross-sectional area (e.g., $A = L_x L_y$ for a cuboid quantization volume); $\varepsilon_s$, the static medium's electric permittivity; and $c_s$, the medium's speed of light. Each mode with wavevector $\mathbf{k}$ and polarization $s$ has natural frequency $\omega_{\mathbf{k}}$, annihilation operator $\hat{a}_{\mathbf{k}}$, and unit polarization vector $\boldsymbol{\varepsilon}_{\mathbf{k},s}$. Naturally, for a linear, isotropic, non-dispersive dielectric medium, the displacement-field operator $\hat{\mathbf{D}}(\mathbf{x}, t)$ is given by

$$\hat{\mathbf{D}}(\mathbf{x}, t) = \varepsilon_s \hat{\mathbf{E}}(\mathbf{x}, t). \tag{S3}$$

If the electromagnetic field corresponds to blackbody radiation, then its first- and second-order field correlations are governed by

$$\begin{aligned} \left\langle \hat{a}_{\mathbf{k},s} \right\rangle =& 0, \\ \left\langle \hat{a}^\dagger_{\mathbf{k},s} \hat{a}_{\mathbf{k}',s'} \right\rangle =& n_{\mathbf{k}} \delta_{\mathbf{k},\mathbf{k}'} \delta_{s,s'}, \\ \left\langle \hat{a}_{\mathbf{k},s} \hat{a}_{\mathbf{k},s} \right\rangle =& 0, \end{aligned} \tag{S4}$$

where $n_{\mathbf{k}}$ is the mean number of photons in mode $\mathbf{k}$ at thermal equilibrium at temperature $T$, i.e.,

$$n_{\mathbf{k}} = \frac{1}{\exp(\hbar\omega_{\mathbf{k}}/k_B T) - 1}. \tag{S5}$$

We aim to represent the initial electromagnetic field in a one-dimensional PTC by a classical random process which yields the same first- and second-order correlations as Eq. (S4) (of course, neglecting non-zero commutation relations, as discussed in Sec. S1). We accomplish this by setting the initial displacement field $D_x(z, t)$ and magnetic field $B_y(z, t)$ as

$$\begin{aligned} D_x(z, 0) =& D_x^{(+)}(z, 0) + \text{c. c.}, \\ B_y(z, 0) =& B_y^{(-)}(z, 0) + \text{c. c.} \end{aligned} \tag{S6}$$

Here, the positive-frequency parts $D_x^{(+)}(z, 0)$ and $B_y^{(+)}(z, 0)$, in analogy to Eq. (S2), are given by

$$\begin{aligned} D_x^{(+)}(z, 0) =& \frac{1}{\sqrt{AL_z}} \sum_k \sqrt{\frac{\hbar\omega_k \varepsilon_s}{2}} a_k \exp(ikz), \\ B_y^{(+)}(z, 0) =& \frac{1}{\sqrt{AL_z}} \sum_k \sqrt{\frac{\hbar\omega_k \mu_0}{2}} \operatorname{sgn}(k) a_k \exp(ikz). \end{aligned} \tag{S7}$$

Now, instead of annihilation operators, the $a_k$ are classical random variables. In analogy to Eq. (S4), we impose that these satisfy

$$\begin{aligned} \langle a_k \rangle =& 0, \\ \langle a_k^* a_{k'} \rangle =& n_k \delta_{k,k'}, \\ \langle a_k a_{k'} \rangle =& 0. \end{aligned} \tag{S8}$$

In analogy to Eq. (S5), $n_k$ is still interpreted as the mean number of photons in thermal equilibrium at temperature $T$, given by

$$n_k = \frac{1}{\exp(\hbar\omega_k/k_B T) - 1}. \tag{S9}$$

We consider a one-dimensional medium with static, non-dispersive phase velocity $c_s$. Hence, the dispersion relation governing $\omega_k$ is

$$\omega_k = c_s |k|. \tag{S10}$$

Of course, the expectation values Eq. (S8) can be realized with several choices of underlying random variables $a_k$. And these would all yield identical first- and second-order field correlations in a PTC. However, a natural choice to satisfy Eq. (S8) is to set the mode amplitudes $a_k$ as

$$a_k = \gamma_k \sqrt{n_k}. \tag{S11}$$

In Eq. (S11), $\gamma_k$ are independent, identically distributed standard complex Gaussian random variables; i.e., their real and imaginary parts are mutually independent normal random variables of variance $1/2$.

The choice Eq. (S11) is further consistent with the expectation that thermal fields follow Gaussian statistics. This expectation follows from (1) the Wigner-function representation discussed in Section S1, (2) the application of the optical equivalence theorem of quantum optics to blackbody radiation [S1], and (3) the observation that thermal fields are formed from the addition of a vast numer of microscopic electromagnetic fields from a thermal emitter [S1, S 7]

We utilize Eq. (S6), Eq. (S7), and Eq. (S11) together as a prescription for the initial conditions of the electromagnetic field for the finite-difference time-domain simulations of thermal radiation in a PTC indicated in the manuscript. We confirm the accuracy of these simulations by comparing the spatial correlations and spectra they yield with those predicted with the correlation-theory formalism of presented in the main manuscript and derived in Sections S3 and S5.

## S3. BLACKBODY RADIATION SPATIAL CORRELATIONS, SPATIAL SPECTRA, AND THEIR NORMALIZATION

In this section, we evaluate the spatial correlations and spatial spectra of the initial classical blackbody radiation of Section S2. We start with the field auto-correlations. Substituting Eq. (S6), Eq. (S7) and Eq. (S8) into the expressions for the field auto-correlations, we find

$$\varepsilon_s^{-1} \langle D_x(z+\Delta z,0)D_x(z,0)\rangle = \mu_0^{-1} \langle B_y(z+\Delta z,0)B_y(z,0)\rangle = \frac{\hbar}{AL_z}\sum_k n_k\omega_k \cos(k\Delta z). \tag{S12}$$

To simplify the right-hand side of Eq. (S12) further, we explicitly consider the limit of large $L_z$ relative to the considered spatial shift $\Delta z$ and the characteristic wavenumber $k_BT/\hbar c_s$. In this limit, we approximate the sum over wavenumbers $k$ as an integral via the prescription

$$\sum_k \to \frac{L_z}{2\pi}\int_{-\infty}^{\infty} \mathrm{d}k\,, \tag{S13}$$

well known in the context of solid-state physics [S8] and modal photonics (both classical [S9, S 10] and quantum [S6, S 11]). The resulting integral for the field auto-correlations can be evaluated in closed form utilizing the Fourier cosine transform pair [S12, S 13]

$$\begin{aligned} \frac{1}{2x^2} - \frac{1}{2}\left(\frac{\pi}{x_0}\right)^2 \operatorname{csch}^2\left(\frac{\pi x}{x_0}\right) &= \int_0^\infty \mathrm{d}k \cos(kx)\frac{k}{\exp(kx_0)-1}, \quad k>0, \\ \frac{k}{\exp(kx_0)-1} &= \frac{2}{\pi}\int_0^\infty \mathrm{d}x \cos(kx)\left[\frac{1}{2x^2} - \frac{1}{2}\left(\frac{\pi}{x_0}\right)^2 \operatorname{csch}^2\left(\frac{\pi x}{x_0}\right)\right]. \end{aligned} \tag{S14}$$

In this way, we find that the fields' spatial auto-correlations in the limit of large quantization length $L_z$ evaluate to [S13, S 14]

$$\varepsilon_s^{-1} \langle D_x(z+\Delta z)D_x(z)\rangle = \mu_0^{-1} \langle B_y(z+\Delta z)B_y(z)\rangle = \frac{\pi k_B^2T^2}{6\hbar c_s A} f(\pi k_B T\Delta z/\hbar c_s). \tag{S15}$$

Here, $f(x)$ the real-analytic function

$$f(x) = \begin{cases} 3x^{-2} - 3\operatorname{csch}^2 x, & x\neq 0, \\ 1, & x=0. \end{cases} \tag{S16}$$

In a parallel way, we evaluate the initial cross-correlation between $D_x(z,0)$ and $B_y(z,0)$. When substituting Eq. (S6), Eq. (S7) and Eq. (S8) into the expression for the cross-correlation, we find

$$\langle D_x(z+\Delta z,0)B_y(z,0)\rangle / \sqrt{\epsilon_s\mu_0} = \frac{\hbar}{AL_z}\sum_k \operatorname{sgn}(k) n_k\omega_k\cos(k\Delta z). \tag{S17}$$

Then, taking the limit of large quantization length $L_z$, Eq. (S13), we obtain

$$\langle D_x(z+\Delta z,0)B_y(z,0)\rangle = 0, \tag{S18}$$

i.e., the fields $D_x(z,0)$ and $B_y(z,0)$ are initially uncorrelated.

Given Eq. (S15), it is then natural to normalize the displacement field $D_x(z,t)$ and magnetic field $B_y(z,t)$ via

$$\begin{pmatrix} \tilde{D}_x(z,t) \\ \tilde{B}_y(z,t) \end{pmatrix} = \sqrt{\frac{6\hbar c_s A}{\pi k_B^2 T^2}} \begin{pmatrix} \tilde{D}_x(z,t)/\sqrt{\varepsilon_s} \\ \tilde{B}_y(z,t)/\sqrt{\mu_0} \end{pmatrix}. \tag{S19}$$

Then, as in the main manuscript, we introduce the abbreviations $\Gamma_j^{(r)}(\Delta z,t)$ $(j = D,B,X)$ for the normalized fields' real-valued spatial correlations

$$\begin{aligned} \Gamma_D^{(r)}(\Delta z,t) &= \langle \tilde{D}_x(z+\Delta z,t)\tilde{D}_x(z,t)\rangle, \\ \Gamma_B^{(r)}(\Delta z,t) &= \langle \tilde{B}_y(z+\Delta z,t)\tilde{B}_y(z,t)\rangle, \\ \Gamma_X^{(r)}(\Delta z,t) &= \langle \tilde{D}_x(z+\Delta z,t)\tilde{B}_y(z,t)\rangle. \end{aligned} \tag{S20}$$

Then, from Eq. (S15) and Eq. (S18), we get

$$\begin{aligned} \Gamma_D^{(r)}(\Delta z,0) =&\Gamma_B^{(r)}(\Delta z,0) = \Gamma_0(\Delta z), \\ \Gamma_X^{(r)}(\Delta z,0) =&0. \end{aligned} \tag{S21}$$

Here, as in the main manuscript, we define the initial normalized auto-correlation $\Gamma_0(\Delta z)$, given by

$$\Gamma_0(\Delta z) = f(\pi k_B \Delta z/\hbar c_s). \tag{S22}$$

In particular, as a direct consequence of Eq. (S16) and Eq. (S22), we have

$$\Gamma_D^{(r)}(0,0) = \Gamma_B^{(r)}(0,0) = 1. \tag{S23}$$

Next, we utilize the Wiener-Khinchine [S1, S 7, S 15] theorem to obtain the spatial spectra $W_j^{(r)}(k,t)$ $(j = D,B,X)$ from the corresponding spatial correlation $\Gamma_j^{(r)}(\Delta z,t)$ via Fourier transformation

$$W_j^{(r)}(k,t) = \int_{-\infty}^{\infty} \mathrm{d}z\, \Gamma_j^{(r)}(\Delta z,t)\exp(-ik\Delta z). \tag{S24}$$

Substituting Eq. (S21) into Eq. (S24), and leveraging again the Fourier cosine transform pair Eq. (S14), we obtain

$$\begin{aligned} W_D^{(r)}(k,t) =&W_B^{(r)}(k,t) = W_0(k,t), \\ W_X^{(r)}(k,0) =&0, \end{aligned} \tag{S25}$$

where the initial spectral density $W_0(k)$ is given by

$$W_0(k) = \frac{6}{\pi}\left(\frac{\hbar c_s}{k_B T}\right)^2 |k| \left[\exp(\hbar c_s |k|/k_B T) - 1\right]^{-1}. \tag{S26}$$

## S4. DEFINITION OF SPATIAL FIGURES OF MERIT

In this section, we define precisely the figures of merit used in the main manuscript to characterize quantitatively the initially blackbody field's spatial correlations $\Gamma_j(\Delta z,t)$ and spatial spectra $W_j(k,t)$ $(j = D,B,X)$. To do so, we first introduce their time-dependent zeroth-order moment $N_j(t)$ via the equivalent integral expressions

$$N_j(t) = \int_{-\infty}^{\infty} \mathrm{d}z \left|\Gamma_j(z,t)\right|^2 = \frac{1}{2\pi}\int_{-\infty}^{\infty} \mathrm{d}k \left|W_j(k,t)\right|^2. \tag{S27}$$

In Eq. (S27), the second equality follows from Parseval's theorem of Fourier analysis. Then, from $\Gamma_j(z,t)$ and $W_j(k,t)$, we may define the time-dependent central wavenumbers $\bar{k}_j(t)$, spatial bandwidths $\Delta k_j(t)$, and coherence lengths $L_j(t)$ $(j = D,B,X)$ via the integrals [S1, S 16, S 17]

$$\begin{aligned} \bar{k}_j(t) &\equiv \left[2\pi N_j(t)\right]^{-1} \int_{-\infty}^{\infty} \mathrm{d}k\, k \left|W_j(k,t)\right|^2, \\ \left[\Delta k_j(t)\right]^2 &\equiv \left[2\pi N_j(t)\right]^{-1} \int_{-\infty}^{\infty} \mathrm{d}k \left[k - \bar{k}_j(t)\right]^2 \left|W_j(k,t)\right|^2, \\ \left[L_j(t)\right]^2 &\equiv \left[N_j(t)\right]^{-1} \int_{-\infty}^{\infty} \mathrm{d}z\, z^2 \left|\Gamma_j(z,t)\right|^2. \end{aligned} \tag{S28}$$

| Figure of merit | Value |
|---|---|
| mean wavenumber, $\bar{k}$ | $0.8522 \times k_B T/\hbar c_s$ |
| uncertainty product, $L\Delta k$ | 1.3378 |
| coherence length, $L$ | $0.2815 \times \hbar c_s/k_B T$ |
| spatial bandwidth, $\Delta k$ | $0.7564 \times k_B T/\hbar c_s$ |

**Table S1.** Spatial figures of merit of the initial blackbody radiation

Because $\Gamma_j(z,t)$ and $W_j(k,t)$ form a Fourier-transform pair, it follows from the definitions Eq. (S28) that the product $L_j \Delta k_j$ is bounded from below by the Fourier-transform limit [S1, S 16]

$$L_j \Delta k_j \geq 1/2. \tag{S29}$$

Furthermore, Eq. (S29) is an equality if and only if $W_j(k,t)$ is a Gaussian function of $k$ (equivalently, $\Gamma_j(z,t)$ is a Gaussian function of $z$).

## S5. ANALYSIS OF MAXWELL'S EQUATIONS IN THE WAVENUMBER DOMAIN

In this section, we derive the time-evolution of the initially blackbody fields' spatial spectra from Maxwell's equations. We consider Maxwell's curl equations for a one-dimensional PTC in the wavenumber domain for a single polarization. Via Fourier transformation over the spatial domain, it follows that these can be written in matrix form as

$$i\frac{\partial}{\partial t}\begin{pmatrix} D_x/\sqrt{\varepsilon_s} \\ B_y/\sqrt{\mu_0} \end{pmatrix} = c_s k \begin{pmatrix} 0 & 1 \\ \eta_r(t) & 0 \end{pmatrix} \begin{pmatrix} D_x/\sqrt{\varepsilon_s} \\ B_y/\sqrt{\mu_0} \end{pmatrix}. \tag{S30}$$

Here, $k$ is the wavenumber, and $\eta_r(t)$ is the time-dependent electric impermittivity relative to its initial static value, i.e.,

$$\eta_r(t) = \varepsilon_s/\varepsilon(t). \tag{S31}$$

As indicated in the main manuscript, Maxwell's equations in the form Eq. (S30) are analogous to those of a harmonic oscillator with time-varying frequency $c_s|k|\sqrt{\eta(t)}$. To see this easily, we take the time-derivative of the first component of Eq. (S30) to obtain

$$\frac{\partial^2 D_x}{\partial t^2} = -\left[c_s|k|\sqrt{\eta(t)}\right]^2 D_x(t). \tag{S32}$$

Given the form of the field first- and second-order moments in Eq. (S11), it is convenient to rewrite Maxwell's equations, Eq. (S30), for the time-dependent amplitudes $a_k(t)$ of the unperturbed medium's propagation modes. By definition, these are the amplitudes of the unperturbed medium's eigenmodes $x^{(\pm)}$. By diagonalization of the square matrix in Eq. (S30), these are

$$x^{(\pm)} = \begin{pmatrix} D_x^{(\pm)}/\sqrt{\varepsilon_s} \\ B_y^{(\pm)}/\sqrt{\mu_0} \end{pmatrix} = \frac{1}{\sqrt{2}}\begin{pmatrix} 1 \\ \pm\mathrm{sgn}(k) \end{pmatrix}, \tag{S33}$$

with eigenfrequencies

$$\omega_k^{(\pm)} = \pm c_s|k|. \tag{S34}$$

In Eq. (S33), $\mathrm{sgn}(\cdot)$ is the sign function. Expressing then Eq. (S30) in the unperturbed mode basis $x^{(\pm)}$, Eq. (S30) becomes

$$i\frac{\partial}{\partial t}\begin{pmatrix} a_k(t) \\ a_{-k}^*(t) \end{pmatrix} = M_k(t)\begin{pmatrix} a_k(t) \\ a_{-k}^*(t) \end{pmatrix}. \tag{S35}$$

In Eq. (S35), we have identified that the amplitude of the unperturbed mode $x^{(-)}$ corresponds to $a_{-k}^*(t)$ as follows from the reality of the spatial-domain fields $D_x(z,t)$ and $B_y(z,t)$ and is consistent with the expression for $x^{(-)}$ in Eq. (S33).

In Eq. (S35), $M_k(t)$ is a square matrix, given by the sum

$$M_k(t) = M_k^{(0)}(t) + V_k(t), \tag{S36}$$

with a diagonal term $M_k^{(0)}(t)$ and off-diagonal term $V_k(t)$. These are correspondingly given by

$$\begin{aligned} M_k^{(0)}(t) =& \omega_k[1 + \Delta\eta(t)/2]\sigma_3, \\ V_k(t) =& i\omega_k \Delta\eta(t)\sigma_2/2. \end{aligned} \tag{S37}$$

In Eq. (S37), $\omega_k \equiv \omega_k^{(+)} = c_s|k|$; $\Delta\eta(t) = [\eta_r(t) - 1]$; and $\sigma_j$ ($j = 0, 1, 2, 3$) stand for the Pauli matrices

$$\sigma_0 = \begin{pmatrix} 1 & 0 \\ 0 & 1 \end{pmatrix}, \quad \sigma_1 = \begin{pmatrix} 0 & 1 \\ 1 & 0 \end{pmatrix}, \quad \sigma_2 = \begin{pmatrix} 0 & -i \\ i & 0 \end{pmatrix}, \quad \sigma_3 = \begin{pmatrix} 1 & 0 \\ 0 & -1 \end{pmatrix}. \tag{S38}$$

To aid the description of the evolution of the spatial spectra $W_j(k,t)$, we introduce the wavenumber-domain time-evolution matrix $U_k(t)$. This is the square matrix which relates the modal amplitude vector $(a_k(t), a_{-k}^*(t))^T$ to its value at the start of PTC modulation $t = 0$, i.e.,

$$\begin{pmatrix} a_k(t) \\ a_{-k}^*(t) \end{pmatrix} = U_k(t) \begin{pmatrix} a_k(0) \\ a_{-k}^*(0) \end{pmatrix}. \tag{S39}$$

Substituting Eq. (S39) into Eq. (S35), it follows that $U_k(t)$ satisfies

$$i\frac{\partial}{\partial t} U_k(t) = M_k(t) U(t). \tag{S40}$$

Furthermore, from Eq. (S39), it has initial condition

$$U_k(0) = 1. \tag{S41}$$

Then, through steps identical to those to compute the initial spatial correlations and spectra, we obtain the time evolution of the spatial spectra. Specifically, we evaluate the spatial correlations, take the limit of infinite $L_z$, and Fourier transform to obtain the spatial spectra. In this way, we obtain

$$\begin{aligned} W_D^{(r)}(k,t) =& W_0(k)\left[S_0(k,t) + S_1(k,t)\right], \\ W_B^{(r)}(k,t) =& W_0(k)\left[S_0(k,t) - S_1(k,t)\right], \\ W_X^{(r)}(k,t) =& -i\mathrm{sgn}(k) W_0(k) S_2(k,t). \end{aligned} \tag{S42}$$

In Eq. (S42), $S_j(k,t)$ ($j = 0, 1, 2, 3$) are the dynamical Bloch-vector components. These are defined in terms of the matrix elements $u_{nm}(k,t)$ of $U_k(t)$ in Eq. (S39) and the Pauli matrices $\sigma_j$ (Eq. (S38)) via

$$S_j(k,t) = \begin{pmatrix} u_{11}^*(k,t) & u_{21}^*(k,t) \end{pmatrix} \sigma_j \begin{pmatrix} u_{11}(k,t) \\ u_{21}(k,t) \end{pmatrix}. \tag{S43}$$

Next, we prove the statements of the main manuscript. These are that the Floquet eigenfrequencies $\omega_k^{(n)}$ ($n = 1, 2$) satisfy

$$\omega_k^{(1)} = -\omega_k^{(2)} \equiv \omega_k \tag{S44}$$

always, and

$$\mathrm{Re}\{\omega_k\} = j\Omega/2 \tag{S45}$$

for some integer $j$ inside the wavenumber bandgaps.

Eq. (S44) follows from applying Abel-Liouville's formula [S18] to the time-evolution operator $U_k(t)$, governed by Eq. (S40). Thus, we obtain

$$\det\{U_k(t)\} = \exp\left[-i\int_0^t \mathrm{d}t' \,\mathrm{tr}\{M_k(t')\}\right] = 1. \tag{S46}$$

The last equality of Eq. (S46) follows from the fact that $M_k(t)$ in Eq. (S36) has vanishing trace for all times $t$. Expressing the determinant $\det\{U_k(t)\}$ in terms of the eigenvalues of $U_k(t)$, then it follows from Eq. (S46) that the product of these eigenvalues equals unity. In particular, by definition, the Floquet eigenfrequencies $\omega_k^{(n)}$ are defined so that the eigenvalues of $U_k(2\pi/\Omega)$ equal $\exp\left[-i2\pi\omega_k^{(n)}/\Omega\right]$. Thus

$$\omega_k^{(1)} + \omega_k^{(2)} = j\Omega \tag{S47}$$

for some integer $j$. But, without loss of generality, we may take $j = 0$, as the Floquet eigenfreqies are meaningful only up to integer multiples of the modulation frequency $\Omega$. Thus, Eq. (S44) follows from Eq. (S47).

We note that Eq. (S44) implies that Floquet modes appear in pairs evolving with opposite frequencies. So it can be interpreted to be a manifestation of discrete time-reversal symmetry of Maxwell's equations in a PTC (discrete over integer multiples of the modulation period $2\pi/\Omega$). But this result follows mathematically because $M_k(t)$ in Eq. (S40) is traceless. Thus, the traceless form of $M_k(t)$ itself can be interpreted to be a form of discrete time-reversal symmetry.

Next, we leverage the pseudo-Hermiticity [S5, S 19, S 20] of Maxwell's equations Eq. (S35) to prove Eq. (S45) inside the wavenumber bandgaps. Maxwell's equations Eq. (S35) are said to be pseudo-Hermitian if there exists an invertible Hermitian matrix $P$ such that [S19]

$$M_k^\dagger(t) = P M_k(t) P^{-1}. \tag{S48}$$

In general, $P$ could be wavenumber- and time-dependent. But it is easy to verify that a constant $P$ satisfying Eq. (S48) exists and is given by

$$P = \sigma_3 \tag{S49}$$

from the Pauli matrices Eq. (S38).

As a consequence of Maxwell's equations for $U_k(t)$, Eq. (S40), and the pseudo-Hermiticity relation Eq. (S48). It follows that

$$\frac{\partial}{\partial t}\left[U_k^\dagger(t) P U_k(t)\right] = 0. \tag{S50}$$

This can be integrated with the initial condition Eq. (S41) for $U_k(t)$ to yield

$$U_k^\dagger(t) P U_k(t) = P. \tag{S51}$$

In other words, the (generally non-positive-definite) metric induced by $P$ is conserved by the time-evolution matrix $U_k(t)$.

By definition, $P$ is invertible (in our case, because $P = \sigma_3$ is invertible). And, from the Abel-Liouville formula Eq. (S46), so is $U_k(t)$. So we may rewrite Eq. (S51) as

$$U_k^\dagger(t) = P U_k^{-1}(t) P^{-1}. \tag{S52}$$

Eq. (S52) then means that $U_k^\dagger(t)$ is similar to $U_k^{-1}(t)$, and thus have the same eigenvalues. Applying this result to $t = 2\pi/\Omega$, this means that the set of eigenvalues $\{\exp(-i2\pi\omega_k/\Omega), \exp(+i2\pi\omega_k/\Omega)\}$ is equal to the set $\{\exp(-i2\pi\omega_k^*/\Omega), \exp(+i2\pi\omega_k^*/\Omega)\}$. This can occur in one of two ways. The first way is if the eigenvalues $\exp(-i2\pi\omega_k/\Omega)$ are of unit magnitude and are thus in a bandpass of the PTC's band-structure. The second way is if

$$\mathrm{Im}\{\exp(-i2\pi\omega_k/\Omega)\} = 0. \tag{S53}$$

Clearly, Eq. (S53) is equivalent to Eq. (S45). Furthermore, inside a wavenumber bandgap, the Floquet eigenvalues $\exp(\pm i2\pi\omega_k/\Omega)$ do not have a modulus of unity, by definition. So pseudo-Hermiticity implies inside such a bandgap, Eq. (S53) and thus Eq. (S45) hold.

As identified in Ref. [S5], the pseudo-Hermiticity property Eq. (S48) is a strong conservation relation of Maxwell's equations, as it implies conservation of the electromagnetic field's phase-space area (or symplectic form) and wavenumber-space Minkowski momentum.

Lastly, owing to the form of Eq. (S49), Eq. (S51) implies that $S_3(k,t)$ in Eq. (S43) is conserved under time evolution. Bloch-space trajectories $(S_1, S_2, S_3)$ are therefore confined to planes of constant $S_3$, in contrast to Hermitian two-level systems, whose trajectories are confined to a Bloch sphere of constant $S_0 = \sqrt{S_1^2 + S_2^2 + S_3^2}$. Because the constant-$S_3$ planes are non-compact, electromagnetic fields in a PTC can grow without bound under resonant time modulation. By contrast, the compact Bloch sphere bounds the amplitude under Hermitian dynamics.

## S6. PSEUDO-HERMITIAN ROTATING-WAVE APPROXIMATION FOR LIGHT IN A PHOTONIC TIME CRYSTAL

The differential equation Eq. (S40) for the time-evolution matrix $U_k(t)$ describes periodically modulated coupling between two detuned modes. Thus, this system is similar to the optically driven coupling between two atomic levels [S21]. Hence, when the modulation is near-resonant, we may do as in the analysis of near-resonantly driven two-level atoms, and introduce a rotating-wave approximation (RWA) to yield an approximate closed-form solution of Eq. (S40) for $U_k(t)$. This is the task of the current section.

To introduce a RWA for $U_k(t)$ in Eq. (S40), we do as in the analysis of near-resonantly driven two-level systems [S21], and first introduce the change of variables

$$U_k(t) = U_k^{(0)}(t)U_k^{(i)}(t), \tag{S54}$$

where $U^{(0)}(t)$ is the exponential matrix

$$U_k^{(0)}(t) = \exp\left[-i\int_0^t \mathrm{d}t'\, M_k^{(0)}(t')\right]. \tag{S55}$$

Substituting Eq. (S54) into Eq. (S40) and Eq. (S41), we find that these three are consistent if and only if $U_k^{(i)}(t, t_0)$ satisfies the differential equation

$$i\frac{\mathrm{d}}{\mathrm{d}t}U_k^{(i)}(t) = V_k^{(i)}(t)U_k^{(i)}(t), \tag{S56}$$

with initial condition

$$U_k^{(i)}(0) = 1. \tag{S57}$$

In Eq. (S56), $V_k^{(i)}(t)$ is given by

$$V_k^{(i)}(t) = \left[U_k^{(0)}(t)\right]^{-1} V_k(t)U_k^{(0)}(t). \tag{S58}$$

Substituting Eq. (S55) into Eq. (S58), we obtain the explicit expression

$$V_k^{(i)}(t) = \frac{i\omega_k\Delta\eta(t)}{2}\left\{\cos\left[2\phi_k(t)\right]\sigma_y + \sin\left[2\phi_k(t)\right]\sigma_x\right\}. \tag{S59}$$

Here, $\phi_k(t)$ is the accumulated phase

$$\phi_k(t) = \omega_k\int_0^t \mathrm{d}t'\left[1 + \Delta\eta(t')/2\right]. \tag{S60}$$

To solve Eq. (S56) approximately, it is convenient to assume an explicit time-dependence for the impermittivity change $\Delta\eta(t)$. Consistent with the main manuscript, we suppose it is sinusoidal, i.e., it is of the form

$$\Delta\eta(t) = \Delta\eta_0\sin(\Omega t + \varphi), \tag{S61}$$

for some modulation amplitude $\Delta\eta_0$, modulation frequency $\Omega$, and modulation phase $\phi$. For the sinusoidal modulation of Eq. (S61), it is convenient to utilize the Jacobi-Anger expansion [S22]:

$$\exp(iz\sin\theta) = \sum_{n=-\infty}^{\infty} J_n(z)\exp(in\theta), \tag{S62}$$

where $J_n(\cdot)$ is the Bessel function of the first kind and of integer order $n$. In this way, we find

$$V_k^{(i)}(t) = \frac{i\omega_k\Delta\eta(t)}{2}\left\{\sum_{n=-\infty}^{\infty} J_n(\Delta\eta_0\omega_k/\Omega)\left[\sigma_y\cos\psi_k^{(n)}(t) + \sigma_x\sin\psi_k^{(n)}(t)\right]\right\} \tag{S63}$$

where

$$\psi_k^{(n)}(t) = (2\omega_k - n\Omega)t - 2\omega_k t_0 - n(\varphi + \pi/2) + \Delta\eta_0\omega_k\cos(\Omega t_0 + \varphi)/\Omega. \tag{S64}$$

To examine the expression in Eq. (S63) for $V_k^{(i)}(t)$, it is useful to recall Eq. (S61) for $\Delta\eta(t)$ and to utilize the trigonometric identities

$$\begin{aligned} \sin x \cos y =& \frac{1}{2}\left[\sin(x-y)+\sin(x+y)\right], \\ \sin x \sin y =& \frac{1}{2}\left[\cos(x-y)-\cos(x+y)\right]. \end{aligned} \tag{S65}$$

In this way, we identify the coefficients multiplying $\sigma_y$ and $\sigma_x$ in each $n$ term of the infinite sum in Eq. (S63) as

$$\frac{1}{2}J_n(\Delta\eta_0\omega_k/\Omega)\left\{\sin\left[\Omega t-\psi_k^{(n)}(t)+\varphi\right]+\sin\left[\Omega t+\psi_k^{(n)}(t)+\varphi\right]\right\}, \tag{S66}$$

and

$$\frac{1}{2}J_n(\Delta\eta_0\omega_k/\Omega)\left\{\cos\left[\Omega t-\psi_k^{(n)}(t)+\varphi\right]-\cos\left[\Omega t+\psi_k^{(n)}(t)+\varphi\right]\right\}, \tag{S67}$$

respectively.

Next, we argue that just a few of the terms in the infinite sum of Eq. (S63) accumulate their contribution and thus actually influence the system's dynamics. Hence, we may just keep these terms and neglect all others. This constitutes the rotating-wave approximation (RWA), popular in the analysis of optical resonances of atomic and molecular transitions. To see this, we note that the oscillation frequency of the terms in Eq. (S66) and Eq. (S67) (i.e., the time-derivative of the argument of the trigonometric functions), is given by

$$\Omega \pm \frac{\mathrm{d}\psi_k^{(n)}}{\mathrm{d}t} = \pm\left[2\omega_k-(n\mp 1)\Omega\right]. \tag{S68}$$

Clearly, Eq. (S68) is closest to zero when

$$\omega_k/\Omega \approx (n\mp 1)/2. \tag{S69}$$

For given values of $\omega_k$ and $\Omega$, Eq. (S69) is satisfied only for a few values of $n$. Hence, indeed only a few of the terms in Eq. (S63) accumulate over time.

In particular, for sufficiently small $|\Delta\eta_0\omega_k/\Omega|$, $J_n(\Delta\eta_0\omega_k/\Omega)\approx J_n(0)=\delta_{n,0}$, where $\delta_{n,m}$ is the Kronecker delta. In this case, Eq. (S69) reduces to

$$\omega_k/\Omega \approx \mp 1/2 \quad \text{for } n=0. \tag{S70}$$

Both $\Omega$ and $\omega_k$ are positive by definition, so Eq. (S70) can be satisfied only via $\omega_k/\Omega\approx 1/2$ rather than $\omega_k/\Omega\approx -1/2$. Thus, for sufficiently small $|\Delta\eta_0\omega_k/\Omega|$, we need retain only the term $n=0$ in Eq. (S63) and the first terms inside the curly brackets in Eq. (S66) and Eq. (S67). Then, in this case, we may simplify Eq. (S63) for $V_k^{(i)}(t)$ to

$$\begin{aligned} V_k^{(i)}(t) \approx& \frac{i\omega_k\Delta\eta_0}{4}J_0(\Delta\eta_0\omega_k/\Omega)\left[\sigma_y\sin\theta_k(t)+\sigma_x\cos\theta_k(t)\right], \\ =& \frac{i\omega_k\Delta\eta_0}{4}J_0(\Delta\eta_0\omega_k/\Omega)\exp\left[-i\theta_k(t)\sigma_z/2\right]\sigma_x\exp\left[i\theta_k(t)\sigma_z/2\right]. \end{aligned} \tag{S71}$$

Here, $\theta_k(t)$ is an abbreviation for

$$\theta_k(t)=\Omega t-\psi_k^{(0)}(t)+\varphi. \tag{S72}$$

Given the simplified form for $V_k^{(i)}(t)$ in Eq. (S71), it is then straightforward to solve the differential equation Eq. (S56) with Eq. (S57) as initial condition. In this manner, we obtain

$$U_k^{(i)}(t,t_0)\approx \exp\left[-i\theta_k(t)\sigma_z/2\right]\exp\left[-i\tilde{M}_k(t-t_0)\right]\exp\left[i\theta_k(t_0)\sigma_z/2\right], \tag{S73}$$

where $\tilde{M}_k$ is given by

$$\tilde{M}_k=\left[\omega_k-\Omega/2\right]\sigma_z+i\omega_k\Delta\eta_0 J_0(\Delta\eta_0\omega_k/\Omega)\sigma_x/4. \tag{S74}$$

The exponentials of Pauli matrices can be evaluated by the general formulas [S23, S 24]

$$\begin{aligned}\exp\left[ia_r\left(\hat{\mathbf{n}}\cdot\vec{\sigma}\right)\right] =&\sigma_0\cos a_r + i\left(\hat{\mathbf{n}}\cdot\vec{\sigma}\right)\sin a_r,\\ \exp\left(\mathbf{a}\cdot\vec{\sigma}\right) =&\sigma_0\cosh\alpha + \left(\mathbf{a}\cdot\vec{\sigma}\right)\frac{\sinh\alpha}{\alpha}.\end{aligned} \tag{S75}$$

These formulas can be verified by direct expansion of the exponentials on the left-hand side and leveraging the property $\sigma_j^2 = \sigma_0$ $(j = 0, 1, 2, 3)$.

Here, $a_r$ is a real-valued coefficient; $\hat{\mathbf{n}}$ $(n_x\hat{\mathbf{x}} + n_y\hat{\mathbf{y}} + n_z\hat{\mathbf{z}})$, a unit vector; and $\mathbf{a}$ $(a_x\hat{\mathbf{x}} + a_y\hat{\mathbf{y}} + a_z\hat{\mathbf{z}})$, a complex-valued vector. $(\hat{\mathbf{n}}\cdot\vec{\sigma})$ and $(\mathbf{a}\cdot\vec{\sigma})$ are the $2\times 2$ matrices

$$\begin{aligned}\hat{\mathbf{n}}\cdot\vec{\sigma} =&n_x\sigma_x + n_y\sigma_y + n_z\sigma_z,\\ \mathbf{a}\cdot\vec{\sigma} =&a_x\sigma_x + a_y\sigma_y + a_z\sigma_z.\end{aligned} \tag{S76}$$

Finally, $\alpha$ is given by

$$\alpha = \sqrt{a_x^2 + a_y^2 + a_z^2}. \tag{S77}$$

Employing these matrix exponentials, the elements $u_{11}(k,t)$ and $u_{21}(k,t)$ of $U_k(t)$ evaluate to

$$\begin{aligned}u_{11}(k,t) =&\exp\left[-i\phi_k(t) - i\theta_k(t)/2\right]\left[\cosh\left(\alpha_k t\right) - i\frac{\Delta\omega_k}{\alpha_k}\sinh\left(\alpha_k t\right)\right],\\ u_{21}(k,t) =&\frac{\kappa_k}{\alpha_k}\exp\left[i\phi_k(t) + \theta_k(t)/2\right]\sinh(\alpha_k t).\end{aligned} \tag{S78}$$

Here, we introduce the abbreviations

$$\begin{aligned}\alpha_k =&\sqrt{\kappa_k^2 - (\Delta\omega_k)^2},\\ \kappa_k =&\omega_k\Delta\eta_0 J_0\left(\Delta\eta_0\omega_k/\Omega\right)/4,\\ \Delta\omega_k =&\omega_k - \Omega/2.\end{aligned} \tag{S79}$$

Having developed the RWA for the pseudo-Hermitian system of light in a PTC, next we leverage this approximation to obtain closed-form expressions for the dynamical Bloch-vector components $S_j(k,t)$ which govern the evolution of the fields' spatial spectra $W_j(k,t)$, as indicated in Eq. (S42). Thus, we substitute Eq. (S78) for the matrix elements $u_{11}(k,t)$ and $u_{21}(k,t)$ into the definition, Eq. (S43). In this manner, we obtain that, to the accuracy of the RWA,

$$\begin{aligned}S_0(k,t) =&\frac{\kappa_k^2}{\alpha_k^2}\cosh(2\alpha_k t) - \frac{(\Delta\omega_k)^2}{\alpha_k^2},\\ S_1(k,t) =&\frac{\kappa_k}{\alpha_k}\left\{\sinh\left(2\alpha_k t\right)\cos[\bar{\rho}_k(t)] - \frac{\Delta\omega_k}{\alpha_k}\left[\cosh(2\alpha_k t) - 1\right]\sin[\bar{\rho}_k(t)]\right\},\\ S_2(k,t) =&\frac{\kappa_k}{\alpha_k}\left\{\sinh\left(2\alpha_k t\right)\sin[\bar{\rho}_k(t)] + \frac{\Delta\omega_k}{\alpha_k}\left[\cosh(2\alpha_k t) - 1\right]\cos[\bar{\rho}_k(t)]\right\},\\ S_3(k,t) =&1.\end{aligned} \tag{S80}$$

Eq. (S80) simplifies further for sufficiently long modulation time $t$ such that the exponentially decaying contributions become negligible; i.e., when $\alpha_k t \gg 1$. In this time regime, the dynamical Bloch-vector components become

$$\begin{pmatrix}S_0(k,t)\\ S_1(k,t)\\ S_2(k,t)\end{pmatrix} = \frac{\kappa_k^2}{2\alpha_k^2}\exp(2\alpha_k t)\begin{pmatrix}1\\ \mathrm{sgn}(\kappa_k)\cos[\rho_k(t)]\\ \mathrm{sgn}(\kappa_k)\sin[\rho_k(t)]\end{pmatrix}. \tag{S81}$$

Here, $\rho_k(t)$ is the time-dependent oscillation phase of the Bloch-vector components. This is given by

$$\rho_k(t) = \chi_k + \Omega t - \Delta\eta_0\omega_k\cos\left(\Omega t + \varphi\right)/\Omega. \tag{S82}$$

## S7. EXPANSION OF BLOCH-VECTOR COMPONENTS AROUND MAIN BANDGAP PEAK

In the main manuscript, we observe the evolution of the initially blackbody spectra towards Gaussian shapes. There, we explain this behavior as a consequence of the PTC's wavenumber bandgap structure. To demonstrate this quantitatively, we leverage the RWA and we analyze the PTC's bandgap to show how this convergence arises. Those are the tasks of this section.

This section is divided into two subsections. In Subsection A, we introduce suitable approximations into the RWA results of Section S6 to derive this convergence to Gaussian spectra. In Subsection B, we leverage Floquet theory by performing an eigenvalue decomposition of the single-period evolution matrix and introducing approximation to them. In this way, we provide a complementary derivation of the convergence towards Gaussian spatial spectra.

### A. Analysis based on the rotating-wave approximation

First, we observe that, by definition, the wavenumber components around the main bandgap peak amplify the most rapidly. Thus, these components dominate the integral expressions for the fields' spatial correlations (from the Wiener-Khinchine theorem) and spatial figures of merit (Section S4). So we may restrict consideration of the wavenumbers $k$ to those around the main bandgap and still obtain accurate estimates for the spatial correlations and spatial figures of merit.

To the accuracy of the RWA, amplification within the PTC's main bandgap is governed by the wavenumber-dependent amplification rate $\alpha_k$ in Eq. (S79). Thus, within this approximation, the main bandgap peak occurs at the wavenumber $k_0$ where $\alpha_k$ attains its maximum. In other words, $k_0$ is such that

$$\alpha'_{k_0} = 0. \tag{S83}$$

In other words, where the derivative of $\alpha_k$ with respect to $k$ vanishes.

If Bessel functions of integer order can be evaluated, as in standard numerical packages, the derivative of $\alpha_k$ with respect to $k$ can be evaluated via elementary methods from Eq. (S79) and the well known recurrence relation for Bessel functions [S22]

$$J_{n-1}(x) - J_{n+1}(x) = 2J'_n(x). \tag{S84}$$

Of course, as a consequence of Eq. (S83), it follows that $\alpha_k$ around $k_0$ can be modeled as a quadratic function of $k$. Knowing that the $k_0$ is a maximum of $\alpha_k$, then we may approximate $\alpha_k$ as

$$\alpha_k \approx \alpha_{k_0} - \frac{1}{2}\left|\alpha''_{k_0}\right|(k-k_0)^2, \tag{S85}$$

for sufficiently small wavenumber deviation $|k-k_0|$. Here, $\alpha''_{k_0}$ stands for the second derivative of $\alpha_k$ with respect to $k$ evaluated at $k=k_0$.

We note that the dynamical Bloch-vector parameters $S_j(k,t)$ (and thus the spatial spectral densities $W_j(k,t)$) are most sensitive to changes in $\alpha_k$ than to other parameters (e.g. $\Delta\omega_k$, $\kappa_k$ and $\rho_k(t)$) in the limit of large modulation time $t$. This is because they depend exponentially only on the product $\alpha_k t$. Thus, we may expand $\alpha_k$ to second order in $k$ around $k_0$, but we need only expand the other parameters to zeroth order in $|k-k_0|$, i.e., take their value as constant and equal to that at $k=k_0$. Then, to this approximation, the non-constant Bloch-vector components $S_j(k,t)$ in Eq. (S81) and the corresponding single-sided spatial densities $W_j(k,t)$ in Eq. (S42) become

$$\begin{aligned}
\begin{pmatrix} S_0(k,t) \\ S_1(k,t) \\ S_2(k,t) \end{pmatrix} &= \frac{\kappa_{k_0}^2}{2\alpha_{k_0}^2} \exp\left[2\alpha_{k_0}t - |\alpha''_{k_0}|(k-k_0)^2 t\right] \begin{pmatrix} 1 \\ \mathrm{sgn}(\kappa_{k_0})\cos[\rho_{k_0}(t)] \\ \mathrm{sgn}(\kappa_{k_0})\sin[\rho_{k_0}(t)] \end{pmatrix}, \\
\begin{pmatrix} W_D(k,t) \\ W_B(k,t) \\ W_X(k,t) \end{pmatrix} &= \frac{\kappa_{k_0}^2}{\alpha_{k_0}^2} W_0(k_0) \exp\left[2\alpha_{k_0}t - |\alpha''_{k_0}|(k-k_0)^2 t\right] \begin{pmatrix} 1+\mathrm{sgn}(\kappa_{k_0})\cos[\rho_{k_0}(t)] \\ 1-\mathrm{sgn}(\kappa_{k_0})\cos[\rho_{k_0}(t)] \\ -i\,\mathrm{sgn}(\kappa_{k_0})\sin[\rho_{k_0}(t)] \end{pmatrix}.
\end{aligned} \tag{S86}$$

We obtain the analytic spatial correlations $\Gamma_j(\Delta z, t)$ corresponding to Eq. (S86) via straightforward

Fourier inversion as

$$\begin{pmatrix} \Gamma_D(\Delta z,t) \\ \Gamma_B(\Delta z,t) \\ \Gamma_X(\Delta z,t) \end{pmatrix} = \frac{\kappa_{k_0}^2 W_0(k_0)}{2\alpha_{k_0}^2\sqrt{\pi|\alpha''_{k_0}|t}} \exp\left[2\alpha_{k_0}t - (\Delta z)^2/(4|\alpha''_{k_0}|t) + ik_0\Delta z\right] \times \begin{pmatrix} 1+\mathrm{sgn}(\kappa_{k_0})\cos[\rho_{k_0}(t)] \\ 1-\mathrm{sgn}(\kappa_{k_0})\cos[\rho_{k_0}(t)] \\ -i\mathrm{sgn}(\kappa_{k_0})\sin[\rho_{k_0}(t)] \end{pmatrix} \tag{S87}$$

Then, in accordance to the main manuscript, we obtain the real-valued spatial correlations $\Gamma_j^{(r)}(\Delta z,t)$ and their Hilbert transforms $\Gamma_j^{(i)}(\Delta z,t)$ as the real part and imaginary part of their corresponding analytic $\Gamma_j(\Delta z,t)$. Thus, we get

$$\begin{pmatrix} \Gamma_D^{(r)}(\Delta z,t) \\ \Gamma_B^{(r)}(\Delta z,t) \\ \Gamma_X^{(r)}(\Delta z,t) \end{pmatrix} = \frac{\kappa_{k_0}^2 W_0(k_0)}{2\alpha_{k_0}^2\sqrt{\pi|\alpha''_{k_0}|t}} \exp\left[2\alpha_{k_0}t - (\Delta z)^2/(4|\alpha''_{k_0}|t)\right] \times \begin{pmatrix} \{1+\mathrm{sgn}(\kappa_{k_0})\cos[\rho_{k_0}(t)]\}\cos(k_0\Delta z) \\ \{1-\mathrm{sgn}(\kappa_{k_0})\cos[\rho_{k_0}(t)]\}\cos(k_0\Delta z) \\ \mathrm{sgn}(\kappa_{k_0})\sin[\rho_{k_0}(t)]\sin(k_0\Delta z) \end{pmatrix}, \tag{S88}$$

and

$$\begin{pmatrix} \Gamma_D^{(i)}(\Delta z,t) \\ \Gamma_B^{(i)}(\Delta z,t) \\ \Gamma_X^{(i)}(\Delta z,t) \end{pmatrix} = \frac{\kappa_{k_0}^2 W_0(k_0)}{2\alpha_{k_0}^2\sqrt{\pi|\alpha''_{k_0}|t}} \exp\left[2\alpha_{k_0}t - (\Delta z)^2/(4|\alpha''_{k_0}|t)\right] \times \begin{pmatrix} \{1+\mathrm{sgn}(\kappa_{k_0})\cos[\rho_{k_0}(t)]\}\sin(k_0\Delta z) \\ \{1-\mathrm{sgn}(\kappa_{k_0})\cos[\rho_{k_0}(t)]\}\sin(k_0\Delta z) \\ -\mathrm{sgn}(\kappa_{k_0})\sin[\rho_{k_0}(t)]\cos(k_0\Delta z) \end{pmatrix}. \tag{S89}$$

And, as a further corollary,

$$\begin{pmatrix} \Gamma_D^{(r)}(0,t) \\ \Gamma_B^{(r)}(0,t) \\ \Gamma_X^{(i)}(0,t) \end{pmatrix} = \frac{\kappa_{k_0}^2 W_0(k_0)}{2\alpha_{k_0}^2\sqrt{\pi|\alpha''_{k_0}|t}} \exp\left(2\alpha_{k_0}t\right) \begin{pmatrix} 1+\mathrm{sgn}(\kappa_{k_0})\cos[\rho_{k_0}(t)] \\ 1-\mathrm{sgn}(\kappa_{k_0})\cos[\rho_{k_0}(t)] \\ -\mathrm{sgn}(\kappa_{k_0})\sin[\rho_{k_0}(t)] \end{pmatrix}. \tag{S90}$$

Eq. (S90) thus yields, to this subsection's approximation, the field variances $\langle D_x^2(t)\rangle = \Gamma_D^{(r)}(0,t)$, $\langle B_y^2(t)\rangle = \Gamma_B^{(r)}(0,t)$, and "effective field cross-variance" $\Gamma_X^{(i)}(0,t)$, as discussed in the main manuscript.

A direct consequence of Eq. (S86) and Eq. (S87) is that, to the accuracy of this subsection's approximations, the spatial figures of merit from Section S4 are straightforward to compute in closed form. Let us denote these as $\bar{k}_{\mathrm{RWA}}$ for the mean wavenumber; $\Delta k_{\mathrm{RWA}}$ for the spatial bandwidth; and $L_{\mathrm{RWA}}$ for the coherence length. Then, these become

$$\begin{aligned} \bar{k}_{\mathrm{RWA}} &= k_0, \\ \Delta k_{\mathrm{RWA}} &= 1/(2\sqrt{|\alpha''_{k_0}|t}), \\ L_{\mathrm{RWA}} &= \sqrt{|\alpha''_{k_0}|t}. \end{aligned} \tag{S91}$$

Naturally, because these approximations for the spatial spectra $W_j(k,t)$ are Gaussian, then

$$L_{\mathrm{RWA}}\Delta k_{\mathrm{RWA}} = 1/2. \tag{S92}$$

### B. Analysis based on Floquet eigenvalue decomposition

As an alternative to the approximation scheme of Subsection A, we propose another one based on the eigenvalue expansion of the wavenumber-domain time-evolution operator $U_k(t)$ after a modulation period $2\pi/\Omega$, i.e., $U_k(2\pi/\Omega)$. To do this, we assume the matrix $U_k(2\pi/\Omega)$ is diagonalizable for all $k$ sufficiently close to the main bandgap peak. Then, we may express $U_k(2\pi/\Omega)$ in terms of its eigenvalue decomposition as [S19, S 20]

$$U_k(2\pi/\Omega) = \sum_{\pm} u_{\pm}(k) x_{\pm}^{(r)}(k) \left[x_{\pm}^{(l)}(k)\right]^{\dagger}. \tag{S93}$$

Because $U_k(2\pi/\Omega)$ is a $2 \times 2$ matrix, it possesses two eigenvalues (possibly repeating for non-single multiplicity). We label one with $(+)$ and one with $(-)$. Then, these eigenvalues are $u_{\pm}(k)$ in Eq. (S93). $x_{\pm}^{(r)}(k)$ are the corresponding right eigenvectors; and $x_{\pm}^{(l)}(k)$, the corresponding left eigenvectors. As conventional, the dagger $\dagger$ stands for the complex-adjoint operation. The eigenvector pairs $x_{\pm}^{(r)}(k)$, $x_{\pm}^{(l)}(k)$ are normalized so that they satisfy

$$\left[x_n^{(l)}(k)\right]^{\dagger} x_m^{(r)}(k) = \delta_{n,m}. \tag{S94}$$

Given Eq. (S93) for $U_k(2\pi/\Omega)$, we may utilize it to evaluate the evolution matrix $U_k(t)$ at integer multiples of the modulation period $2\pi/\Omega$. Because the PTC's medium properties are, by definition, periodic with period $2\pi/\Omega$, then by mathematical induction

$$U_k(2\pi n/\Omega) = U_k^n(2\pi/\Omega). \tag{S95}$$

Hence, substituting Eq. (S93) into Eq. (S95) and leveraging Eq. (S94), we obtain

$$U_k(2\pi n/\Omega) = \sum_{\pm} u_{\pm}^n(k) x_{\pm}^{(r)}(k) \left[x_{\pm}^{(l)}(k)\right]^{\dagger} \approx u_{+}^n(k) x_{+}^{(r)}(k) \left[x_{+}^{(l)}(k)\right]^{\dagger}. \tag{S96}$$

In the last expression of Eq. (S96), we assume $n$ is large enough so that the magnitude of the smaller eigenvalue of $U_k(2\pi n/\Omega)$, $|u_{-}^n(k)|$, is negligible compared to that of the larger one, $|u_{+}^n(k)|$.

Next, we introduce the Floquet amplification rate $\alpha_F(k)$ through the defining relation

$$u_{+}(k) = \exp\left[2\pi\alpha_F(k)/\Omega\right]. \tag{S97}$$

In other words, for each wavenumber $k$, $\alpha_F(k)$ is the effective rate of increase of the temporal mode $x_{+}^{(r)}(k)$ which grows geometrically with increasing number of modulation periods. Substituting Eq. (S97) into Eq. (S96), The latter becomes

$$U_k(t) \approx \exp\left[\alpha_F(k)t\right] x_{+}^{(r)}(k) \left[x_{+}^{(l)}(k)\right]^{\dagger}. \tag{S98}$$

Of course, Eq. (S98) is accurate only for modulation times $t$ equal to integer multiples of the modulation period $2\pi/\Omega$. Nonetheless, we extend its use to times $t$ beyond this discrete set of integer multiples of $2\pi/\Omega$. In this extension, we understand that Eq. (S98) describes only the field dynamics averaged over a modulation period $2\pi/\Omega$.

As in Subsection A, we observe that the wavenumber components around the main bandgap peak amplify the most rapidly. Thus, these components dominate the fields' spatial correlations and spatial figures of merit. So we may expand $\alpha_F(k)$ with respect to $k$ around its maximum and still obtain accurate estimates for the spatial correlations and spatial figures of merit. Thus, we define $k_1$ as the wavenumber maximizing $\alpha_F(k)$ with respect to $k$, and hence satisfying

$$\alpha_F'(k_1) = 0. \tag{S99}$$

Thus, in Eq. (S96), we expand $\alpha_F(k)$ in $k$ around $k_1$ to its first non-constant order, i.e., second order. In contrast, we expand the eigenvectors $x_{+}^{(r)}(k)$ and $x_{+}^{(l)}(k)$ around $k_1$ only to zeroth order. Explicitly, we approximate

$$\begin{aligned} \alpha_F(k) &\approx a_F(k_1) - \frac{1}{2}|\alpha_F''(k_1)|(k-k_1)^2, \\ x_{+}^{(r)}(k) &\approx x_{+}^{(r)}(k_1), \\ x_{+}^{(l)}(k) &\approx x_{+}^{(l)}(k_1). \end{aligned} \tag{S100}$$

As in Subsection A, we justify expanding $\alpha_F(k)$ to second order in $(k-k_0)$, but $x_+^{(r)}(k)$ and $x_+^{(l)}(k)$ only to zeroth order by noting that, as the modulation time $t$ increases, the sensitivity of $U_k(t)$ to changes in $\alpha_F(k)$ grows exponentially; but that to changes in $x_+^{(r)}(k)$ and $x_+^{(l)}(k)$ remains constant.

Substituting then the expansions Eq. (S100) into Eq. (S98), the latter becomes

$$U_k(t) \approx \exp\left[\alpha_F(k_1)t - |\alpha_F''(k_1)|(k-k_1)^2 t/2\right] x_+^{(r)}(k_1) \left[x_+^{(l)}(k_1)\right]^\dagger . \tag{S101}$$

Given the simplified wavenumber- and time-dependence of $U_k(t)$ in Eq. (S101) it is then straightforward to substitute Eq. (S101) into Eq. (S42) and evaluate the corresponding spatial spectra $W_j(k,t)$, spatial correlations $\Gamma_j(\Delta z, t)$, and the spatial figures of merit (Section S4). It readily follows that the latter figures of merit are given by

$$\begin{aligned} \bar{k}_F =& k_1, \\ \Delta k_F =& 1/(2\sqrt{|\alpha_F''(k_1)|t}), \\ L_F =& \sqrt{|\alpha_F''(k_1)|t}. \end{aligned} \tag{S102}$$

Again, because the resulting approximations for the spatial spectra $W_j(k,t)$ are Gaussian, then

$$L_F \Delta k_F = 1/2. \tag{S103}$$

## S8. BLOCH-VECTOR COMPONENTS OF MATRICES

In this section, we state the relation between the entries of the cross-correlation matrix $\bar{\Gamma}(\Delta z, t)$ and the cross-spectral density matrix $\bar{W}(k,t)$ and the Bloch-vector components, $S_j^{(\Gamma)}(\Delta z, t)$ and $S_j^{(W)}(k,t)$, for their eigenvector with largest eigenvalue. These Bloch-vector components are defined more precisely in the main manuscript. It is known in the spinor-vector formalism [S24, S 25] that these Bloch-vector components for a general $2\times 2$ Hermitian matrix are directly expressible in terms of the entries of their corresponding matrix. Applying this result directly to the matrices $\bar{\Gamma}(\Delta z, t)$ and $\bar{W}(k,t)$, it is straightforward to show that the non-zero components of these are given by

$$\begin{aligned} S_1^{(\Gamma)}(\Delta z, t) =& \frac{\Gamma_X^{(r)}(\Delta z, t)}{N_\Gamma(\Delta z, t)}, \\ S_3^{(\Gamma)}(\Delta z, t) =& \frac{\Gamma_D^{(r)}(\Delta z, t) - \Gamma_B^{(r)}(\Delta z, t)}{N_\Gamma(\Delta z, t)}, \\ S_2^{(W)}(\Delta z, t) =& -\frac{\mathrm{Im}\{W_X(k,t)\}}{N_W(k,t)}, \\ S_3^{(W)}(\Delta z, t) =& \frac{W_D(k,t) - W_B(k,t)}{N_W(k,t)}. \end{aligned} \tag{S104}$$

The normalization factors $N_m \geq 0$ ($m = \Gamma, W$) are defined to satisfy $\sum_{n=1}^{3}\left[S_n^{(m)}\right]^2 = 1$ for all times $t$. The Bloch component $S_2^{(\Gamma)}(\Delta z, t)$ vanishes because $\bar{\Gamma}(\Delta z, t)$ is purely real; and $S_1^{(W)}(k,t)$ vanishes because the off-diagonal entries of $\bar{W}(k,t)$ are purely imaginary.

Next, we briefly discuss the physical meaning of these Bloch-vector parameters. From Eq. (S104), $S_3$ is a normalized measure of the imbalance between the diagonal entries of the corresponding matrix. If $|S_3| \sim 1$, the one of the two entries of both eigenvectors of the corresponding Hermitian matrix have almost negligible norm. On the other hand, $\sqrt{S_1^2 + S_2^2}$ is then a measure of how mixed the eigenvectors of the corresponding Hermitian matrix are. The magnitude of $S_1$ relative to $S_2$ measures how close to in-phase/out-of-phase the two entries in the eigenvectors are, as opposed to in-quadrature.



\end{document}